\documentclass[10pt, conference, compsocconf]{IEEEtran}
\IEEEoverridecommandlockouts
\usepackage{cite}
\usepackage{amsmath,amssymb,amsfonts}
\usepackage{algorithmic}
\usepackage{graphicx}
\usepackage{textcomp}
\usepackage{xcolor}
\def\BibTeX{{\rm B\kern-.05em{\sc i\kern-.025em b}\kern-.08em
    T\kern-.1667em\lower.7ex\hbox{E}\kern-.125emX}
}

\usepackage[utf8]{inputenc}
\usepackage[american]{babel}
\usepackage[hidelinks]{hyperref}
\usepackage[capitalize]{cleveref}
\usepackage{mathtools}
\usepackage{subcaption}
\usepackage{booktabs}
\usepackage{fancyhdr}

\usepackage{glossaries}

\newacronym{ap}{AP}{Access Point}
\newacronym{cfr}{CFR}{channel frequency response}
\newacronym{cnn}{CNN}{convolutional neural network}
\newacronym{csi}{CSI}{channel state information}
\newacronym{cv}{CV}{computer vision}
\newacronym{dl}{DL}{deep learning}
\newacronym{har}{HAR}{human activity recognition}
\newacronym{lan}{LAN}{local-area network}
\newacronym{lstm}{LSTM}{long short-term memory}
\newacronym{mimo}{MIMO}{multiple-input multiple-output}
\newacronym{nic}{NIC}{network interface card}
\newacronym{ofdm}{OFDM}{orthogonal frequency-division multiplexing}
\newacronym{ofdma}{OFDMA}{orthogonal frequency-division multiple access}
\newacronym{phy}{PHY}{Physical Layer}
\newacronym{sdr}{SDR}{software-defined radio}
\newacronym{siso}{SISO}{single-input single-output}
\newacronym{sta}{STA}{station}
\newacronym{wlan}{WLAN}{wireless local-area network}

\newcommand*\wifi{\mbox{Wi-Fi}}
\newtheorem{property}{Property}

\begin{document}

\title{\huge Exposing the CSI: A Systematic Investigation of CSI-based Wi-Fi Sensing Capabilities and Limitations}

\author{\IEEEauthorblockN{Marco Cominelli*$^{\dagger}$, Francesco Gringoli*$^{\dagger}$ and Francesco Restuccia$^{\ddagger}$\vspace{-0.3cm}}\\
\IEEEauthorblockA{
* National Inter-University Consortium for Telecommunications~(CNIT), Italy\\
$\dagger$ Department of Information Engineering, University of Brescia, Italy\\
$\ddagger$ Institute for the Wireless Internet of Things, Northeastern University, United States
}
\normalsize E-mail: \{marco.cominelli, francesco.gringoli\}@unibs.it, frestuc@northeastern.edu
}

\maketitle

\thispagestyle{fancy}
\renewcommand{\headrulewidth}{0pt}
\fancyhead{}
\fancyhead[C]{\Large Accepted for publication in Proceedings of IEEE PerCom 2023}

\begin{abstract}
Thanks to the ubiquitous deployment of \wifi{} hotspots, \gls{csi}-based \wifi{} sensing can unleash game-changing applications in many fields, such as healthcare, security, and entertainment.
However, despite one decade of active research on \wifi{} sensing, most existing work only considers legacy IEEE 802.11n devices, often in particular and strictly-controlled environments.
Worse yet, there is a fundamental lack of understanding of the impact on \gls{csi}-based sensing of modern \wifi{} features, such as 160-MHz bandwidth, \gls{mimo} transmissions, and increased spectral resolution in IEEE 802.11ax (\wifi{}~6).
This work aims to shed light on the impact of \wifi{}~6 features on the sensing performance and to create a benchmark for future research on \wifi{} sensing.
To this end, we perform an extensive \gls{csi} data collection campaign involving 3 individuals, 3 environments, and 12 activities, using \wifi{}~6 signals.
An anonymized ground truth obtained through video recording accompanies our 80-GB dataset, which contains almost two hours of \gls{csi} data from three collectors.
We leverage our dataset to dissect the performance of a state-of-the-art sensing framework across different environments and individuals.
Our key findings suggest that (i) \gls{mimo} transmissions and higher spectral resolution might be more beneficial than larger bandwidth for sensing applications; (ii) there is a pressing need to standardize research on \wifi{} sensing because the path towards a truly environment-independent framework is still uncertain.
To ease the experiments' replicability and address the current lack of \wifi{}~6 \gls{csi} datasets, we release our 80-GB dataset to the community.
\end{abstract}

\glsresetall

\section{Introduction}
\label{sec:introduction}

It was the late 1990s when Vic Hayes first introduced the concept of an international standard for wireless networking, which would later become IEEE~802.11 in 1997, also known as \wifi{}.
Initially, \wifi{} was supposed to be a low-rate replacement of Ethernet and supported up to 2~Mbit/s connectivity.
Today, \wifi{} is undeniably one of the most pervasive wireless technologies ever invented, providing up to 14~Gbit/s connectivity in homes, offices, university campuses, parks, airports, and shopping malls, to name a few \cite{khorov2018tutorial}.
To attest to this explosive growth, the number of \wifi{} 6 hotspots is expected to increase 13x by 2023 \cite{cisco2020cisco}, while the global \wifi{} economy is poised to increase by \$1.5T in two years \cite{WiFiAlliance}.

\begin{figure}
  \includegraphics[width=\columnwidth]{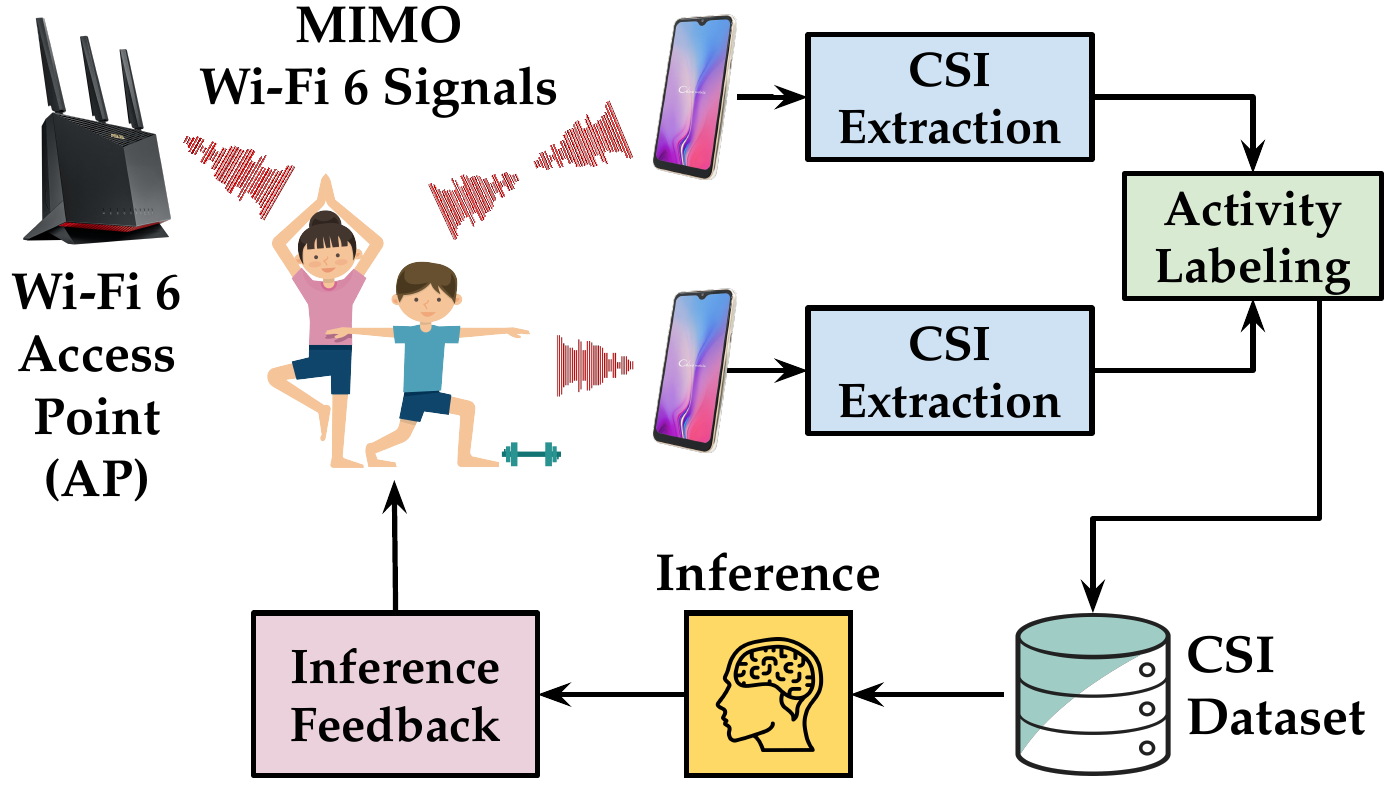}
  \caption{Example of how a CSI-based Wi-Fi sensing framework might be used for \acrshort{har} applications.}
  \label{fig:wireless-sensing}
\end{figure}

Besides connectivity, \wifi{} signals can also be used as fine-grained ``radar'' pulses to \textit{sense} what happens in the surroundings~\cite{li2020passive}.
Among the most popular techniques there is \gls{csi}-based sensing, which aims at capturing small-scale changes in the \gls{cfr} produced by the presence of obstacles located between \wifi{} nodes, such as phase differences~\cite{wang2017phasebeat} or Doppler shifts~\cite{meneghello2022}.
The subtle changes in the \gls{cfr} can be estimated through pilot symbols~\cite{ieee-80211ax} and used to classify phenomena of interest, such as \gls{har}  \cite{korany2020teaching,wei2019real,zheng2019zero,xiao2021onefi}, person detection and identification \cite{korany2021,zeng2016wiwho,soltanaghaei2020human}, human pose tracking \cite{ren2021winect,ren2021tracking}, baggage tracking \cite{shi2021environment} and respiration monitoring \cite{zeng2020multisense}, among many others \cite{ma2019wifi,ma2021location,nirmal2021deep}.
\Cref{fig:wireless-sensing} shows an example in the context of \gls{har}, where a \gls{csi}-based sensing framework tracks the users' fitness activities and sends appropriate recommendations to adjust the daily activity levels.
However, even if research in this field is thriving, we have identified a few key issues that, in the long run, might prevent \gls{csi}-based sensing from unleashing its full potential.

\subsection*{\textbf{Key Research Issues}}

\emph{Legacy CSI signals.}
Since the seminal ``See through walls with WiFi!'' paper in 2013~\cite{seethroughwalls}, \gls{csi}-based sensing has been investigated by a deluge of research work (see \cref{sec:related-work}).
However, most of the current art is still based on legacy 802.11n platforms that fail to capture critical innovations of newer standards, such as wider channels (up to 160-MHz bandwidth) and more subcarriers (up to 2048 per antenna per spatial stream).
To put things in perspective, the increase in \gls{csi} dimensionality between 802.11n and 802.11ax is about ten times the increase in the number of pixels between a PAL-format video and an 8k-resolution video.

\emph{Lack of a full-fledged investigation.}
Despite the unprecedented sensing possibilities provided by the newer versions of \wifi{}, a systematic investigation into \textit{if, when, and how} modern features may improve---or impair---\gls{csi}-based sensing performance is missing.
For example, current systems might even suffer from broader bandwidths or \gls{mimo} transmissions, as they might introduce more noise or inter-stream interference in the sensing framework.
Achieving this fundamental understanding becomes crucial in light of the recent standardization efforts of the IEEE 802.11bf Task Group (TG), founded in September 2020 and aimed at  enabling \wifi{} sensing in common routers~\cite{TGbfPAR}.

\emph{Scarcity of \acrshort{csi} datasets.}
As of October 2022, there were very few \gls{csi} datasets available to the community (see Section \ref{sec:related-work}). Collecting large quantities of \gls{csi} data and preparing ready-to-use labeled datasets is a cumbersome endeavor requiring a lot of effort and specific software tools~\cite{csitool2011,athcsi2015,nexmoncsi2019,picoscenes2021,axcsi2021}, as discussed in Section \ref{sec:related-work}.
Unfortunately, this causes some authors to refrain from sharing their own \gls{csi} datasets, a practice that hinders breakthrough innovation and---more critically---the replicability of the research outcomes.
This does not happen in more mature fields such as in \gls{cv},  where massive datasets like COCO~\cite{lin2014microsoft} or ImageNet~\cite{russakovsky2015imagenet} have been available to the research community for 8+ years.

\subsection*{\textbf{Summary of Novel Contributions}}

$\bullet$ We present the first full-fledged investigation into the capabilities and limitations of \gls{csi}-based \wifi{} sensing with 802.11ax (\wifi{}~6) signals, henceforth referred to as \textit{\gls{csi} sensing} for brevity.
Given its significance, we cast our study into the context of \gls{har}, the most explored problem in \gls{csi} sensing so far \cite{ma2019wifi}.
Specifically, we consider a \gls{har} problem targeting twelve different activities (see \cref{tab:activities}).
To collect the data for our analysis, we set up an extensive data collection campaign involving three subjects in three different indoor environments, as detailed in \cref{sec:data-collection}.
Our experiments involve one \wifi{}~6 transmitter and three \gls{csi} collectors that extract up to 8192 data points per \gls{csi}, i.e., more than 250x the number of data points extracted with legacy 802.11n systems (usually 30 data points per \gls{csi}).
Moreover, \gls{csi} data are synchronized to an anonymized video ground truth, constructed by recording the activities with a webcam, and anonymized by building a 3D model of the subject performing the activity.
\textit{To the best of our knowledge, this is the first \gls{csi} dataset available for \wifi{}~6 systems, and the first introducing video-based ground truth generation}.\smallskip

$\bullet$ We leverage our dataset to dissect the impact of available bandwidth, subcarrier spacing, and number of antennas on \gls{csi} sensing performance.
To this end, we test the performance of one of the most recent frameworks, SHARP~\cite{meneghello2022}, while changing such parameters.
We also evaluate the generalization performance of the system on different subjects and environments.
Our key findings are that more antennas and finer subcarrier spacing greatly improve the sensing performance, while broader bandwidths lead only to a marginal improvement.
\textit{Besides informing the wireless research community, we expect our results will inform ongoing standardization efforts in IEEE 802.11bf}.\smallskip

$\bullet$ To allow full replicability and for benchmarking purposes, we release to the research community our labeled \gls{csi} dataset, including the anonymized ground truth videos, for a total of about 80~GB of data.\footnote{See \url{https://github.com/ansresearch/exposing-the-csi}}
\textit{Beyond eliciting ``pure'' \gls{csi} sensing, we expect our dataset will be used to perform novel research at the intersection of \gls{cv} and \gls{csi} sensing}.

\section{Related Work}
\label{sec:related-work}

\begin{table}
    \centering
    \caption{Non-exhaustive list of some of the most recent (last 3 years) works on \acrshort{csi}-based \wifi{} sensing (A = Activity Recognition, G = Gesture Recognition).}
    \begin{tabular}{lcccl} 
        \toprule
        Name (Year) & 802.11 & Task & Approach \\
        \midrule
        WiAR \cite{guo2019wiar} (2019) & n & A & Traditional ML \\
        Widar 3.0 \cite{zheng2019zero} (2019) & n & G & CNN + GRUs \\
        RF-Net \cite{rfnet2020} (2020) & n   & A & LSTM + Meta-Learning \\
        OneFi \cite{xiao2021onefi} (2021) & n   & G & One-Shot Learning  \\
        ReWiS \cite{bahadori2022rewis} (2022) & ac & A& Few-Shots Learning \\
        SHARP \cite{meneghello2022} (2022) &  ac & A & CNN (Inception v4)\\
        \bottomrule
    \end{tabular}
    \label{tab:summary_rel_work}
\end{table}

The field of \gls{csi} sensing sparked with the development of \gls{csi} extraction tools, i.e., software that enabled researchers to \gls{csi} data from commercial \wifi{} devices.
Then, over the last decade, \gls{csi} sensing has been proposed for a wide variety of applications; among the most compelling, we mention person detection and identification~\cite{korany2021,zeng2016wiwho,soltanaghaei2020human}, localization~\cite{moustafa19}, baggage tracking~\cite{shi2021environment}, and human pose tracking~\cite{ren2021winect,ren2021tracking}, with most of the previous research activities focused on human activity and gesture recognition~\cite{abdelnasser19,korany2020teaching,wei2019real,zheng2019zero,xiao2021onefi}. The applications are so compelling that the IEEE 802.11bf Task Group is currently working to define the appropriate modifications to existing Wi-Fi standards to enhance sensing capabilities through 802.11-compliant devices \cite{chen2022wifi,meneghello2022toward,restuccia2021ieee}.
In \cref{tab:summary_rel_work}, we summarize some of the most recent approaches (i.e., less than three years old) that are most related to the work conducted in this paper, which we will briefly review in the following.
We refer the reader to \cite{ma2019wifi,ma2021location,nirmal2021deep} for excellent survey papers on the topic.
In \cref{tab:csi-extraction-tools}, we report a list of the most common \gls{csi} extraction tools available today.

\begin{table}
  \centering
  \caption{Most common \acrshort{csi} extraction tools available for commercial hardware.}
  \begin{tabular}{lcr}
    \toprule
    CSI Extractor & 802.11 & CSI points \\
    \midrule
    802.11n CSI Tool~\cite{csitool2011} & n & 30 \\
    Atheros CSI Tool~\cite{athcsi2015} & n & 56 \\
    Nexmon CSI~\cite{nexmoncsi2019} & ac & up to 4096 \\
    PicoScenes~\cite{picoscenes2021} & ax & up to 7968 \\
    AX-CSI~\cite{axcsi2021} & ax & up to 32768 \\
    \bottomrule
  \end{tabular}
  \label{tab:csi-extraction-tools}
\end{table}

WiAR~\cite{guo2019wiar} is a framework for \gls{har} based on the 802.11n \gls{csi} Tool.
The WiAR dataset contains sixteen activities from three subjects in three different environments, but it is collected using legacy 802.11n signals.
The authors tested the accuracy of seven traditional \gls{dl} algorithms, showing up to 90\% accuracy with 16 activities.
Based on the same \gls{csi} extraction tool, RF-Net~\cite{rfnet2020} presented a \gls{har} framework using \gls{lstm} layers.
However, the approach is evaluated on a dataset with only four activities, making the findings not widely generalizable.
Regarding gesture recognition, Widar~3.0~\cite{zheng2019zero} and  OneFi~\cite{xiao2021onefi} collected datasets with 802.11n signals containing 6 and 40 different gestures, respectively.
Widar~3.0 derives a Body Velocity Profile (BVP) quantity from the \gls{csi} and introduces gated recurrent units (GRUs) in the framework architecture to improve generalization and sensing performance.
On the other hand, OneFi proposes an approach based on One-Shot Learning (OSL) to recognize previously unseen gestures with only one (or few) labeled samples.
However, while the Widar~3.0 dataset is currently available online, the authors of OneFi replied when contacted that they are not planning to release their dataset to the community, which ultimately makes the findings not replicable.

Even if Nexmon~CSI was released in 2019, very few works have considered \gls{har} in the context of 802.11ac.
Among them, SHARP~\cite{meneghello2022} proposes a technique based on  Doppler shift estimation through the \gls{csi}.
The idea is that Doppler shifts are not affected by environment-specific static objects.
The authors claim that the proposed technique is person- and environment-independent, with an average accuracy of 95\%.
Finally, ReWiS~\cite{bahadori2022rewis} is a framework for \gls{csi}-based \gls{har} using Few-Shots Learning (FSL) to generalize across different environments.
Although code and dataset were made available by the authors, the number of activities considered is only four.

\textbf{Discussion.}~We have identified a series of problems with the public \gls{csi} datasets that are available today: \textbf{(i)} different works consider different activities; \textbf{(ii)} many works consider a limited amount of activities (often less than 8); \textbf{(iii)} many papers only test their implementation with their own dataset; \textbf{(iv)} some public datasets still use the Intel 5300 chipset, which includes only 30 subcarriers per \gls{csi} and is now discontinued by Intel; \textbf{(v)} there is no way to evaluate ``intuitively'' how the activity has been performed; \textbf{(vi)} there is no ground truth to correlate an exact moment during the activity execution with the corresponding \gls{csi}.
After much exploration, we could not find a research paper investigating rigorously the limits and the trade-offs imposed by different \wifi{} standards.
Many interesting questions arise when studying how different features of the \wifi{} signals affect \gls{csi} sensing.
For example, does the sensing accuracy improve with a broader channel bandwidth? And what happens if we consider the reduced subcarrier spacing of 802.11ax?
To our knowledge, this paper is the first to address these questions.

\section{Background on CSI Sensing }
\label{sec:csi-sensing-background}

For the purpose of communication, the \gls{csi} is a crucial element of \wifi{} because it is one of the simplest ways to properly equalize wide-band communication signals received on frequency-selective channels.
In this section, we briefly review the background underpinning \gls{csi} measurements.

\subsection{Background on Wi-Fi PHY Layer}

\Gls{ofdm} has been at the core of \wifi{} since IEEE~802.11a/n\cite{ieee-80211ax}, either at 2.4~GHz, 5~GHz, and now even 6~GHz.
Although \wifi{}~6 introduces \gls{ofdma} to multiplex the data across multiple users, the \gls{csi} is computed starting from the training symbols in the \wifi{} preamble, which are modulated using \gls{ofdm}.
Therefore, here we summarize the basic notions of OFDM, and refer the interest reader to \cite{khorov2018tutorial} for an exhaustive tutorial on \gls{ofdma} on 802.11ax. 

\Gls{ofdm} transceivers subdivide a wide-band channel into $W$ orthogonal subchannels depending on (i) the available bandwidth and (ii) the frequency spacing between adjacent subchannels.
In \cref{tab:ieee-standard-comparison}, we compare the maximum channel bandwidth and the minimum subcarrier spacing allowed by different \wifi{} versions.
For example, the maximum number of \gls{ofdm} subcarriers supported by 802.11n devices is $W=128$ on a 40-MHz channel, while 802.11ax systems can fit $W=2048$ subcarriers into a 160-MHz channel. 

\begin{table}
  \centering
  \caption{Maximum channel bandwidth and \acrshort{ofdm} subcarrier spacing comparison between different Wi-Fi standards.}
  \begin{tabular}{lrr}
    \toprule
    IEEE Standard & Max Bandwidth & Subcarrier Spacing\\
    \midrule
    802.11n (2009) &  40 MHz & 312.5 kHz \\
    802.11ac (2013) & 160 MHz & 312.5 kHz \\
    802.11ax (2021) & 160 MHz & 78.125 kHz \\
    \bottomrule
  \end{tabular}
  \label{tab:ieee-standard-comparison}
\end{table}

\Gls{ofdm} systems map groups of transmit bits onto $W'$ symbols, and each symbol is used to modulate a different \gls{ofdm} subcarrier using traditional modulation techniques, e.g., \mbox{64-QAM}.
In general, $W'<W$ because some subcarriers are reserved to carry \textit{pilot} tones or forced to zero (like the guard bands), but we will drop the distinction between $W$ and $W'$ in the following, as all the subcarriers are then transmitted within a single \gls{ofdm} symbol.
The $k$\textsuperscript{th} \gls{ofdm} symbol $x_k$, transmitted at time $t \in \left[ \, kT, \, (k+1)T \, \right)$ with duration $T$, can be described by:
\begin{equation}
  x_k(t) = \sum_{w=1}^{W} a_{w,k} \exp \left[ j 2 \pi \left( f_c + \frac{f_w}{T} \right) t \right] \, ,
  \label{eq:ofdm}
\end{equation}
where $a_{w,k} \in \mathbb{C} $ is the constellation point modulating the $w$\textsuperscript{th} subcarrier of the $k$\textsuperscript{th} symbol, $f_w$ is the base-band frequency of the $w$\textsuperscript{th} subcarrier, and $f_c$ is the central frequency of the \wifi{} channel. 

\begin{figure}
  \includegraphics[width=\columnwidth]{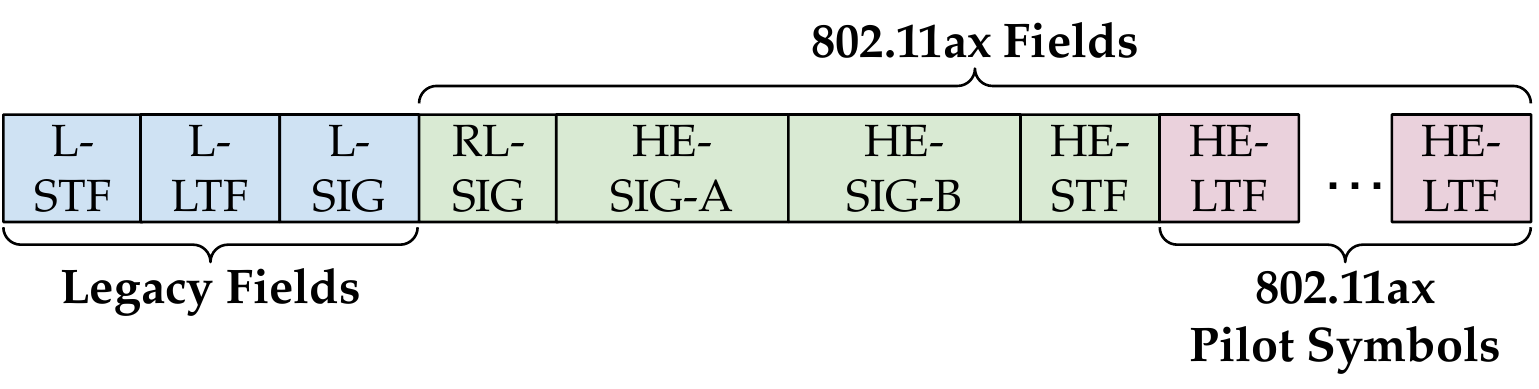}
  \caption{PHY frame in 802.11ax.}
  \label{fig:wifi-frame}
\end{figure}

\Cref{fig:wifi-frame} shows the preamble of an 802.11ax \gls{phy}  frame. At the beginning, training fields are included for retro-compatibility with legacy Wi-Fi standards. Specifically, the legacy short training field (L-STF) and legacy long training field (L-LTF) are used for frame synchronization, while the legacy signal field (L-SIG) is used  to compute the frame duration.
To simplify the 802.11ax frame detection, the HE part of the preamble starts with a repetition of the L-SIG field called RL-SIG, followed by HE-SIG-A.
The latter indicates modulation and coding scheme (MCS), bandwidth, and the number of spatial streams used, among others, while HE-SIG-B contains the \gls{ofdma} \textit{resource allocation} and \textit{resource block}-specific information.
Finally, the \gls{phy} frame preamble is followed by a set of HE-LTF pilot symbols, which are used to estimate the \gls{cfr}, as explained below.

\subsection{Mathematical definition of Channel State Information}

Let us consider an \gls{ofdm} communication system with $W$ subcarriers.
In the following, we focus on one specific symbol, so that we can drop the index $k$ and simplify the notation.
Given one \gls{ofdm} symbol, we can express \cref{eq:ofdm} as the product of a vector of modulating coefficients $\mathbf{x} = [ a_1, \dots, a_W ]$ with a symbol-independent vector containing all the complex exponentials.
Then, the relation between the transmit signal $\mathbf{x} \in \mathbb{C}^W$ and the received one $\mathbf{y} \in \mathbb{C}^W$ is given by:
\begin{equation}
    \mathbf{y} = \mathbf{H} \circ \mathbf{x} \, ,
    \label{eq:communication}
\end{equation}
where $\mathbf{H} \in \mathbb{C}^W$ is the frequency response of the wide-band wireless channel and $\circ$ is the Hadamard product, i.e., the element-wise product.

Since the \gls{ofdm} channel is in general frequency-selective, the receiver needs to estimate its effect in order to properly equalize the received signal.
The \gls{csi} $\mathcal{H} \in \mathbb{C}^W$ is a \textit{quantized} approximation of the channel response $\mathbf{H}$ that is computed by the receiver using the HE-LTF symbols of the \wifi{} preamble in \cref{fig:wifi-frame} as described in the specification~\cite{ieee-80211ax}.
Using the \gls{csi}, the receiver can invert \cref{eq:communication} to equalize the received signal and eventually recover the original transmit signal:
\begin{equation}
    \mathbf{x} \simeq \mathcal{H}^{-1} \circ \mathbf{y} \, .
    \label{eq:equalization}
\end{equation}

We omit spatial multiplexing in our discussion because we will not deal with multi-stream transmissions in our experiments.
However, we will consider the case when the receiver has $N > 1$ antennas.
In this case, it is easy to generalize \cref{eq:communication,eq:equalization} by computing $\mathbf{y}_i$ for every receiving antennas $i$ from 1 to $N$.
This means that from a single transmission we can measure $N$ different \gls{csi} $\mathcal{H}_i$ simultaneously.

\begin{figure}
    \centering
    \includegraphics[width=.98\columnwidth]{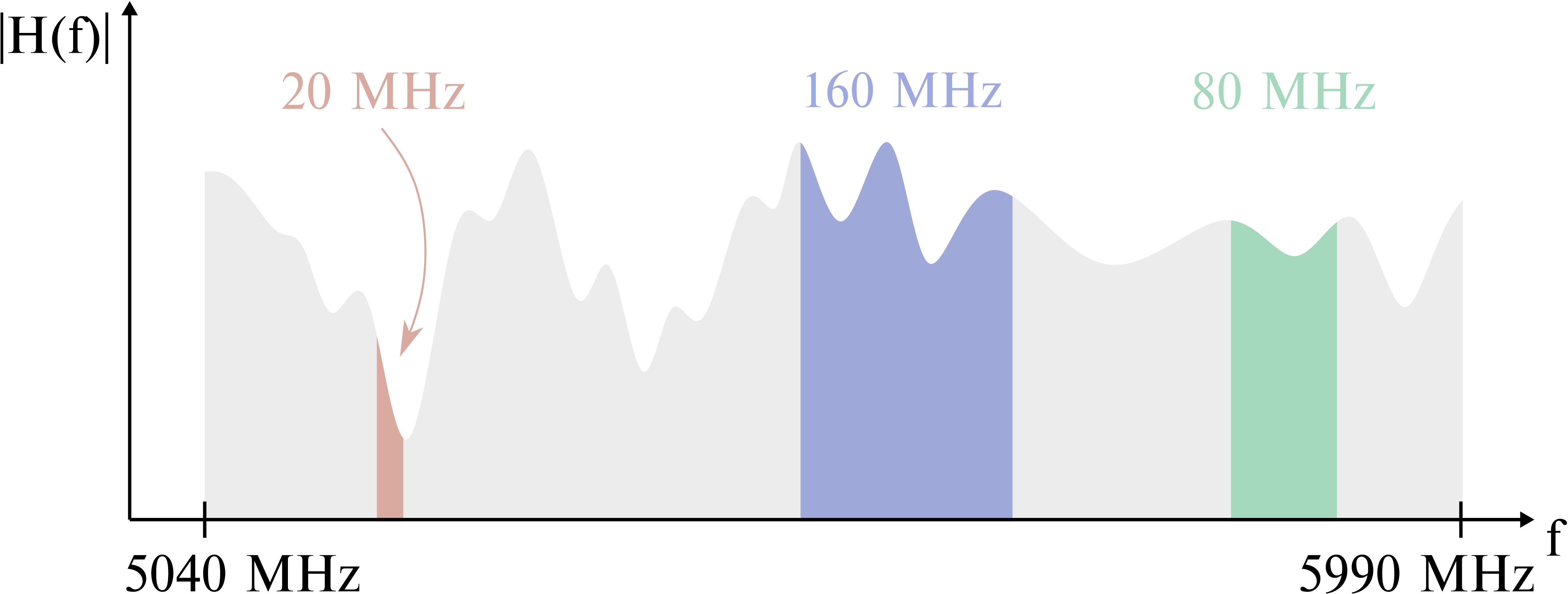}
    \caption{Overview of the 5 GHz ISM band with representation of different bandwidths to scale. Many state-of-the-art works still consider 20-MHz bands only.}
    \label{fig:spectrum5ghz}
\end{figure}

\textbf{Discussion.}
It is clear that the number of antennas, the channel bandwidth and the subcarrier spacing are all key parameters for the resulting \gls{csi}.
In \cref{fig:spectrum5ghz}, we visually represent to scale the bandwidth difference between 20-MHz and 160-MHz systems, and we recall that most bibliography only considers the smallest one.
However, beside the intuition that ``more data, the better," there are more profound implications that should motivate us to consider larger bandwidths and higher frequency resolution in \gls{csi} sensing applications.
The specific interference pattern associated to the \gls{csi} is generated by the superposition at the receiver of several copies of the signal (multi-path propagation). The  properties of the wireless channel measured in the frequency domain by the \gls{csi} can be equivalently described in the time domain through the channel's impulse response.
We recall that one of the basic properties of the Fourier transform asserts that, given a signal $x(t)$ function of time $t$ and its Fourier transform $X(f)$ as a function of frequency $f$, then
\begin{equation}
    x(t - \tau) \xleftrightarrow{\mathcal{F}} X(f) \cdot \exp \left( -j 2 \pi f \tau \right) \, ,
    \label{eq:delay-transform}
\end{equation}
where $\mathcal{F}$ represents the Fourier transform operator and $\tau$ is a time delay.
From \cref{eq:delay-transform} and the linearity of the Fourier transform itself, it is easy to show that receiving multiple copies of the \wifi{} signal at different $\tau$'s (e.g., due to multi-path) turns into a combination of complex exponentials in the frequency domain.
From this, we can derive that a broader \gls{csi} bandwidth helps discriminating lower-frequency exponentials, i.e., paths that arrived close to the first one, while higher spectral resolution can help detecting higher-frequency exponentials, i.e., multi-path components with larger delays.

\section{Methodology}
\label{sec:methodology}

In this work, we want to propose a framework for collecting and processing the \gls{csi} to evaluate the capabilities and limitations of \wifi{} sensing systematically.
Before describing the setup of the experiments, we must define upfront the variables that we will consider in our analysis.
In our investigation, we propose to evaluate the performance of \wifi{} sensing systems while changing three key properties of the \gls{csi}: i) the bandwidth, which can range from 20~MHz to 160~MHz; ii) the subcarrier spacing (either 312.5~kHz for 802.11ac  or 78.125~kHz for 802.11ax); iii) the number of receiving antennas (from one to four).

\subsection{Hardware Selection and Testbed Setup}
\label{ss:testbed-setup}

The fundamental building block of any \gls{csi}-based sensing framework is the \gls{csi} collector, i.e., the device used for measuring the \gls{csi}.
In this work, we opt for the AX-CSI tool because it supports the latest 802.11ax standard with up to 160~MHz bandwidth and 4x4~\gls{mimo} transmissions~\cite{axcsi2021}.

Our testbed consists of four Asus RT-AX86U routers, connected together to form a \gls{lan} and controlled by a host laptop.
During the experiments, one router generates dummy \wifi{} traffic at a constant rate using the injection features of AX-CSI.
The other three routers---which we will also call \textit{monitors}---perform the \textit{sensing}, i.e., they extract the corresponding \gls{csi} from the received dummy \wifi{} frames.
We deploy this testbed in different environments to perform the data collection campaign described in detail in the following sections.

Notice that \gls{csi} data grows very quickly when working with multi-antenna 160-MHz 802.11ax transmissions; hence, storing the \gls{csi} on every monitor can become expensive and inefficient. Thus, the extracted \gls{csi} data is forwarded to the host laptop through the \gls{lan}.
The raw data are then stored as a \texttt{pcap} file on an external storage device.

\subsection{Preliminary Analysis}\label{sec:preliminary-analysis}

Before collecting the dataset, we validate two crucial properties of the \gls{csi}.
Albeit intuitive, their analytical derivation is beyond the scope of this paper.
In this section, we prove that such intuitive properties hold in our measurements by using experimental evidence.

\begin{figure}
    \centering
    \includegraphics[width=\columnwidth]{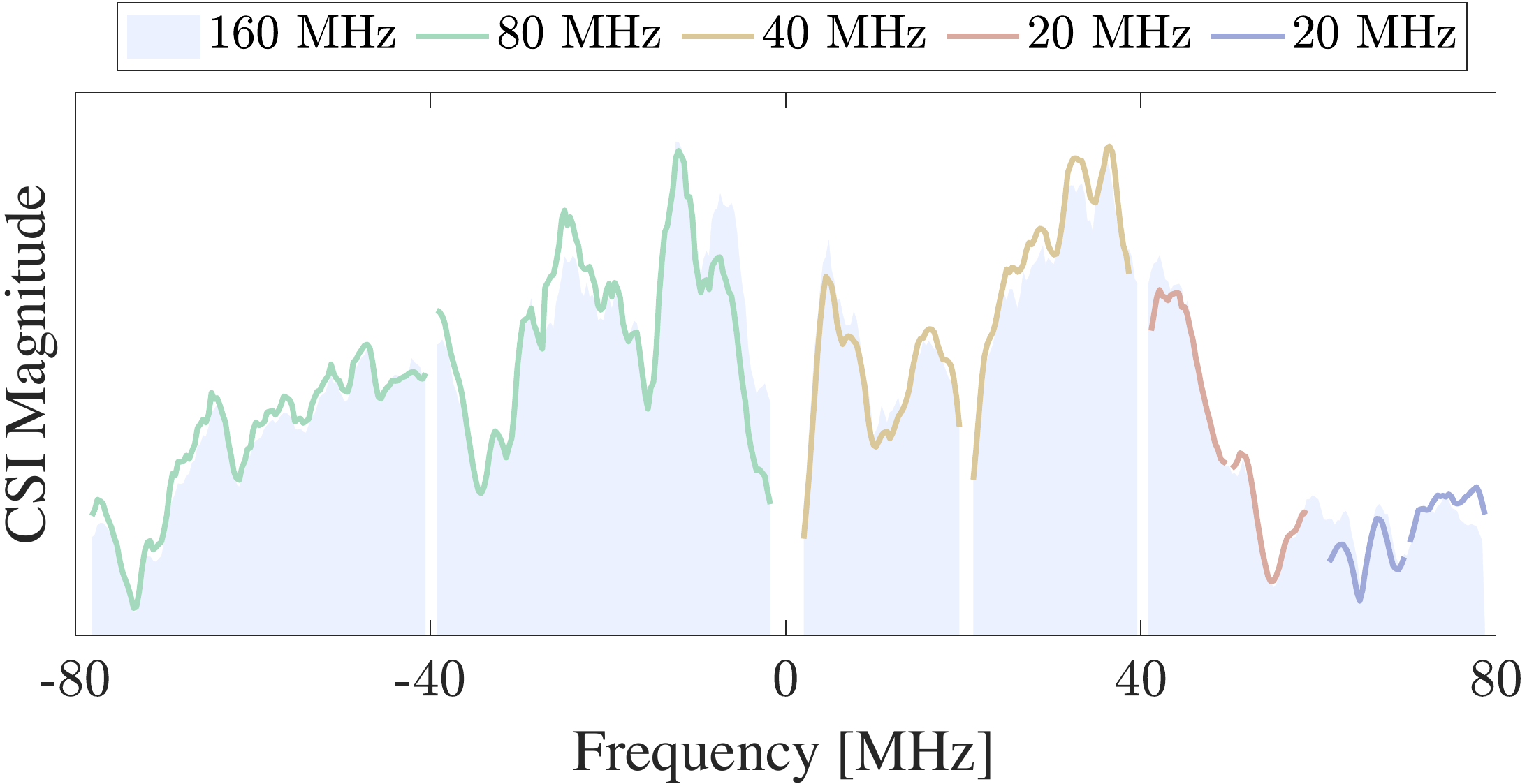}
    \caption{The \acrshort{csi} of different sub-channels represents different portions of the same channel response.}
    \label{fig:band-composition}
\end{figure}

\begin{property}
{\it The \gls{csi} of a wide-band channel is equivalent to the ensemble of \gls{csi} of its narrow-band sub-channels.}
\end{property}
We have discussed in \cref{sec:csi-sensing-background} how the \gls{csi} is a quantized approximation of the \gls{cfr}.
It is clear that no matter the bandwidth of the \gls{csi} being measured, the underlying \gls{cfr} should not change.
In \cref{fig:band-composition}, we show that the \gls{csi} measured for a 160-MHz channel overlaps perfectly (apart from some inevitable noise) with the \gls{csi} of its narrow-band sub-channels, thus proving this point. This result allows us to perform only one experiment in every scenario with the broadest bandwidth, i.e., $160$~MHz.
The \gls{csi} of narrower channels can then be easily derived by considering just the subset of \gls{ofdm} subcarriers corresponding to that narrow-band channels.

\begin{figure}
    \centering
    \includegraphics[width=0.98\columnwidth]{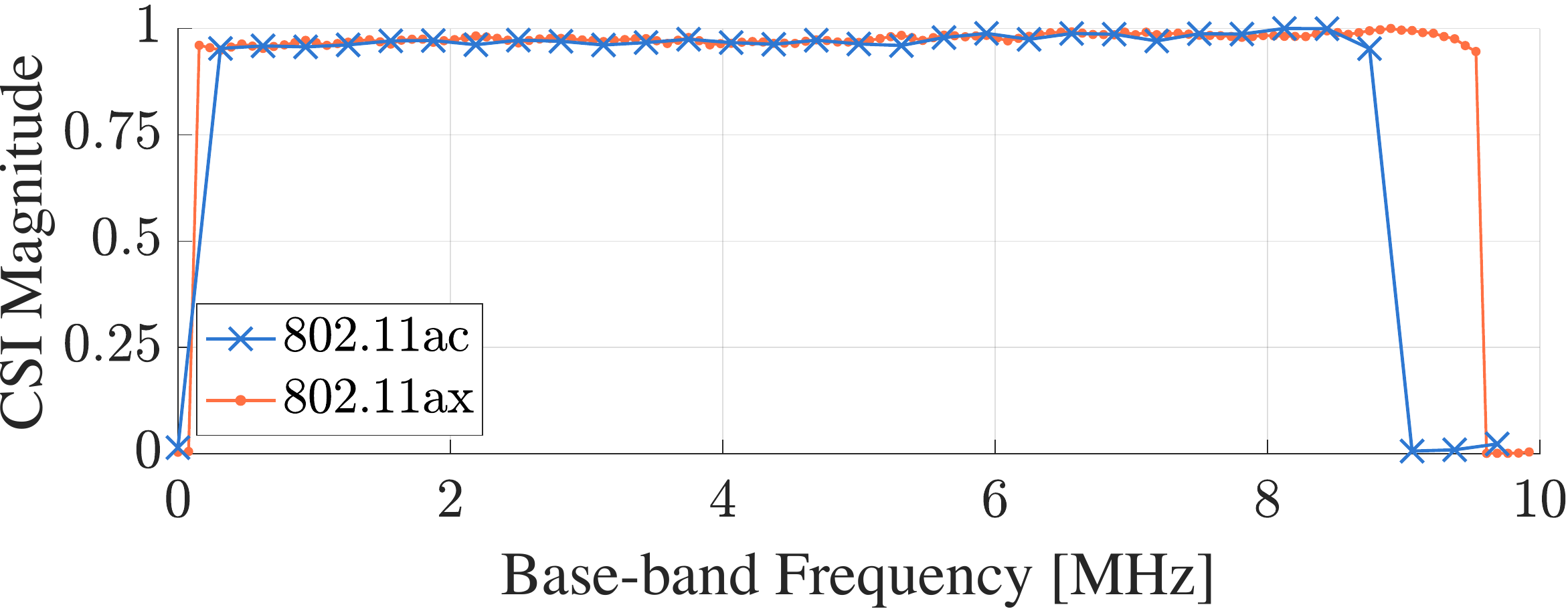}
    \caption{Detail of the \acrshort{csi} magnitude (normalized) measured with 802.11ac and 802.11ax systems over the same channel.}
    \label{fig:band-equivalence}
\end{figure}

\begin{property}
{\it The \gls{csi} measured by 802.11ax and 802.11ac systems represents the same underlying \gls{cfr}.}
\end{property}
With reference to \cref{tab:ieee-standard-comparison}, the smaller subcarrier spacing enables 802.11ax systems to detect higher frequency components in the \gls{csi}, but the general behavior is indeed equivalent.
We prove this in \cref{fig:band-equivalence}, showing a detail of the \gls{csi} measured over the same channel by 802.11ac and 802.11ax systems. This result allows us to run the experiments with 802.11ax transmissions and then do a downsampling operation on the \gls{csi} when we need to consider the \gls{csi} of 802.11ac systems.

\subsection{Data Collection}
\label{sec:data-collection}

\begin{figure*}
    \centering
    \begin{subfigure}{0.32\textwidth}
        \centering
        \includegraphics[width=0.98\columnwidth]{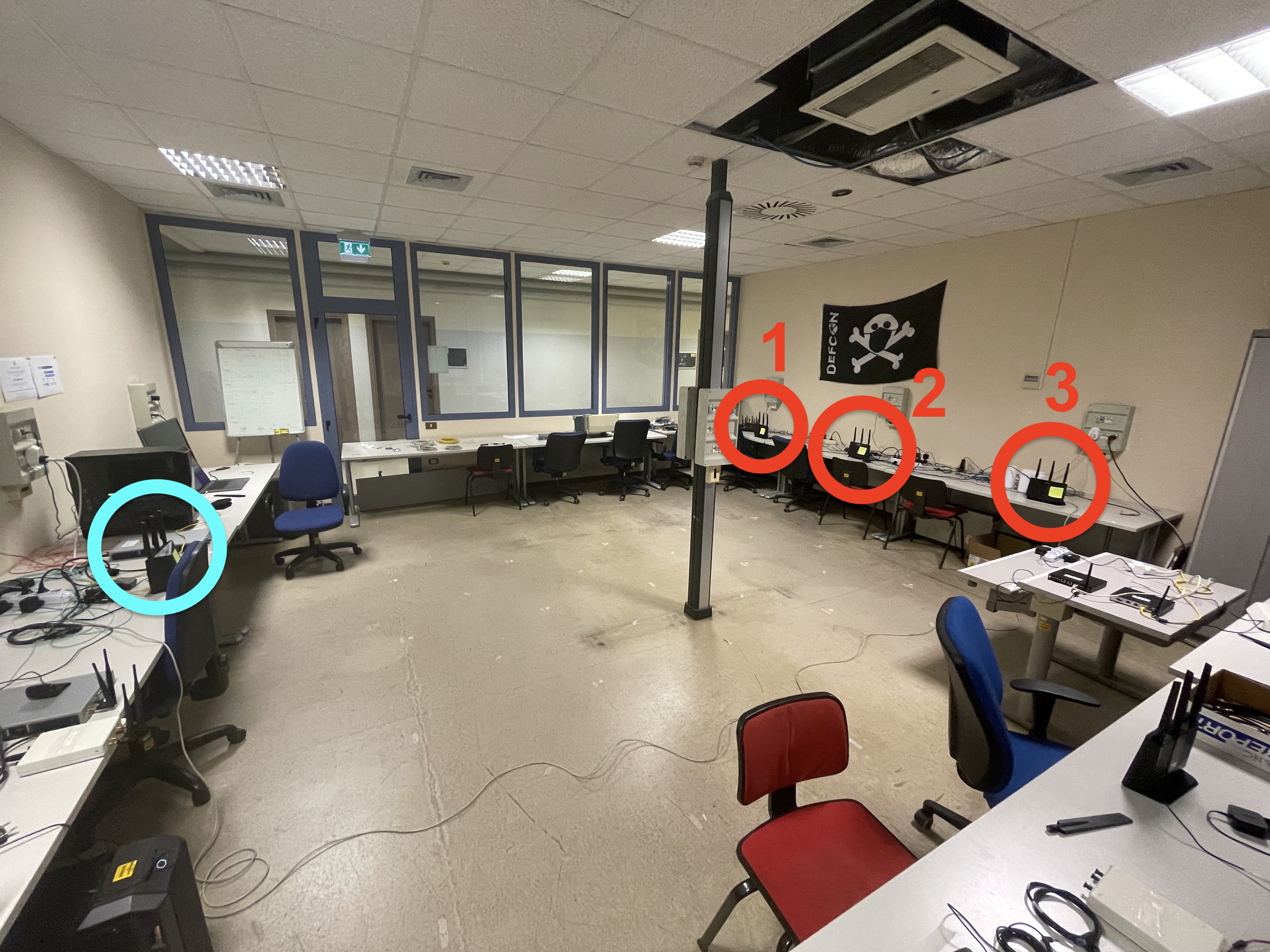}
        \caption{Lab}
        \label{fig:env:lab}
    \end{subfigure}
    \begin{subfigure}{0.32\textwidth}
        \centering
        \includegraphics[width=0.98\columnwidth]{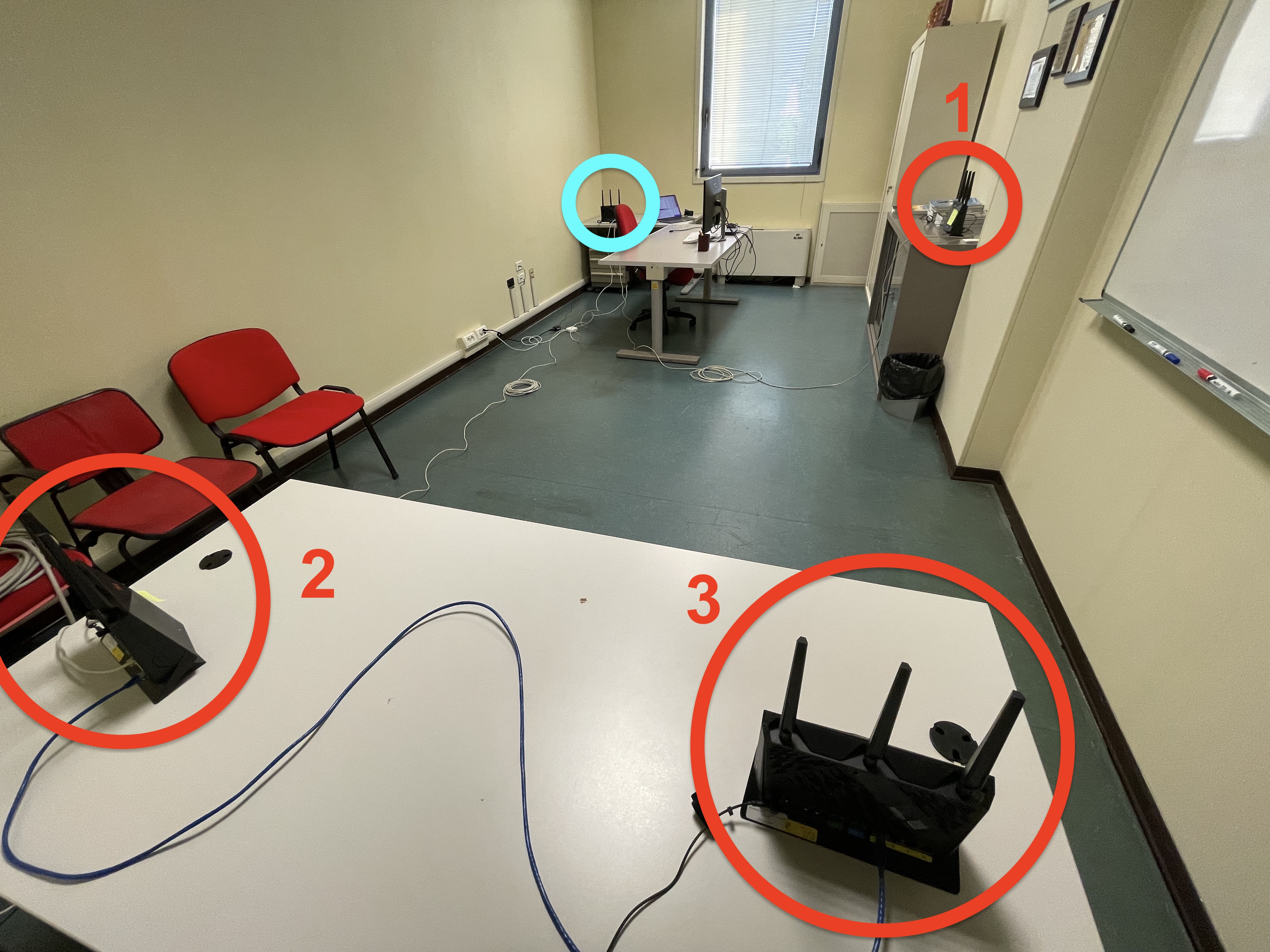}
        \caption{Office}
        \label{fig:env:office}
    \end{subfigure}
    \begin{subfigure}{0.32\textwidth}
        \centering
        \includegraphics[width=0.98\columnwidth]{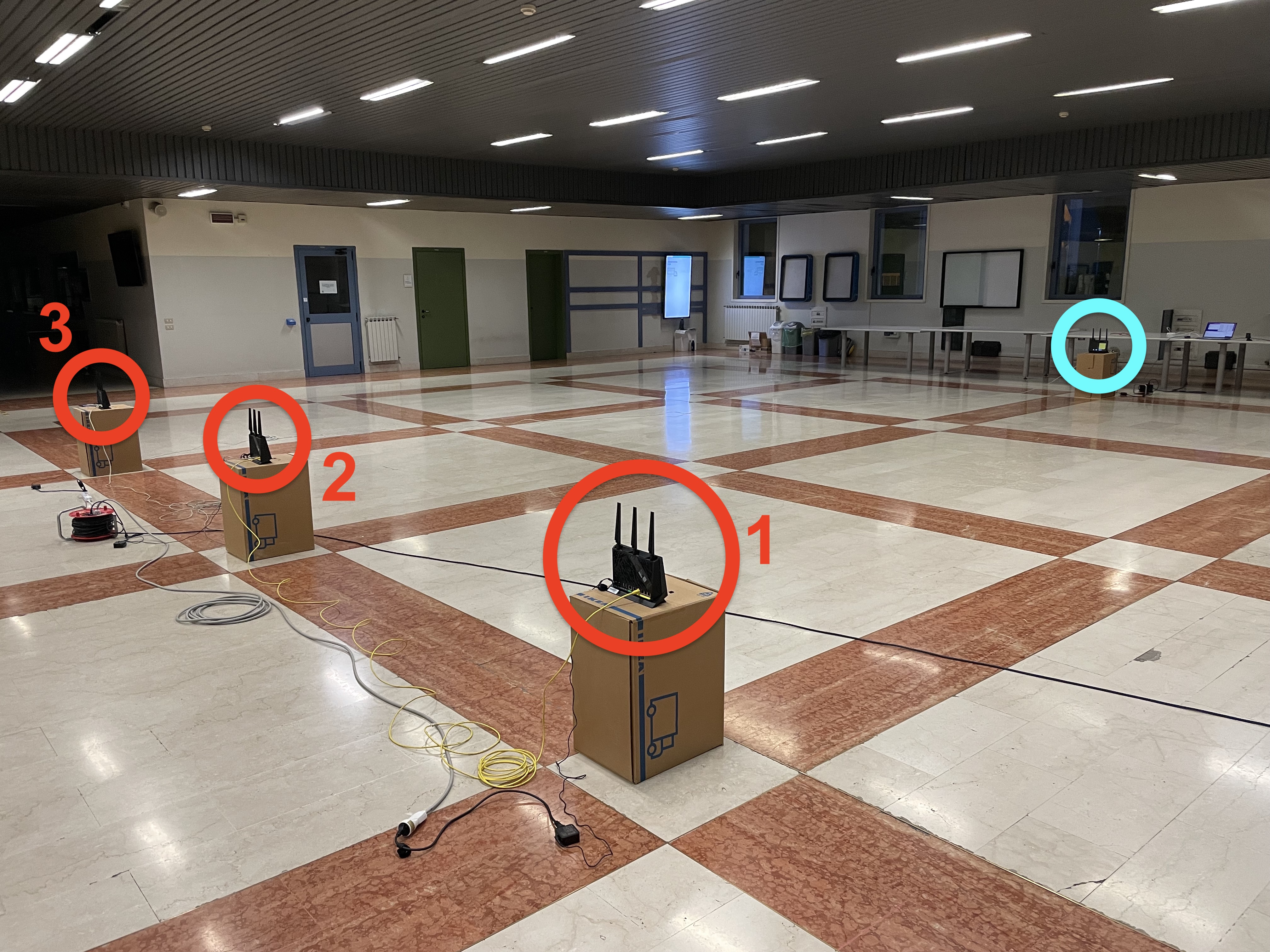}
        \caption{Hall}
        \label{fig:env:hall}
    \end{subfigure}
    \caption{Photographs of the environments considered for the data collection campaign. The transmitter and the three receivers are highlighted with blue and red circles, respectively.}
    \label{fig:environments}
\end{figure*}

Our main experimental activity consists of collecting \gls{csi} data in 7 different experiments, or \textit{scenarios}.
We report the complete list of the considered scenarios in \cref{tab:scenarios} and discuss the rationale behind our specific choice.
First, we would like to validate the property of a sensing framework to work well over time \textit{on the same person in the same environment}; hence, we collect this type of data in S1, S4, and S5.
Note that S1 and S4 are collected a few hours apart during the same day, while S5 data have been collected after one day.
Second, we want to verify that a sensing framework can generalize to  \textit{different people in the same environment}; this type of insight can be given by the results obtained in scenarios S1, S2, and S3, which are three experiments we ran in the lab with three different candidates.
Third, we want to investigate how the sensing framework generalizes to \textit{different environments with the same person}; the scenarios used to benchmark this property are scenarios S5, S6, and S7.
Overall, we believe that these seven scenarios represent the most simple yet complete set of cases to benchmark the most desirable properties of any \gls{csi}-based sensing framework.

\begin{table}
    \centering
    \caption{Overview of the experimental scenarios.}
    \begin{tabular}{cccc}
        \toprule
         Scenario & Candidate & Environment & Day \# \\
        \midrule
         S1 & A & Lab & 1\\
         S2 & B & Lab & 1\\
         S3 & C & Lab & 1\\
         S4 & A & Lab & 1\\
         S5 & A & Lab & 2\\
         S6 & A & Office & 2\\
         S7 & A & Hall & 2\\
        \bottomrule
    \end{tabular}
    \label{tab:scenarios}
\end{table}

In our experiments we consider three different environments, shown in \cref{fig:environments}.
The three chosen environments are representative of different general ``layouts'' of an indoor space:
the \textit{lab} (\cref{fig:env:lab}) represents a medium-sized indoor space, i.e., a sort of ``average'' room in which \gls{csi} sensing can be performed;
the \textit{office} (\cref{fig:env:office}) represents a smaller, more cluttered space rich of multipath reflections;
finally, the \textit{hall} (\cref{fig:env:hall}) represents a larger space with less multipath, except for the one coming from the floor and the ceiling.
Following the discussion in \cref{ss:testbed-setup}, we deploy three monitors and one transmitter in every environment, respectively identified by red and blue circles in \cref{fig:environments}.
We think it is important to remark that scenarios S1 to S5 refer to the same physical environment and that we moved neither the receivers nor the transmitter for the entire duration of the corresponding experiments.
Thus, it is safe to assume that eventual modifications in the channel are due to natural variations in the environmental conditions and do not depend on the position of the devices.

In every scenario, we invite the candidate to perform 12 different activities, and we let every monitor collect 80~s of \gls{csi} data for each activity.
Ultimately, our experimental data collection campaign totaled two hours of \gls{csi} data for each monitor.
The activities performed by the candidate are either static (sitting, standing) or dynamic (walking, waving hands); some of them involve mild physical exercise (jumping, running), but we have selected activities that are compatible with a general office setting.
The complete list of the activities is reported in \cref{tab:activities}.

\begin{table}
  \centering
  \caption{List of the 12 activities considered in our data collection campaign.}
  \begin{tabular}{l@{\quad}ll@{\quad}ll@{\quad}ll@{\quad}l}
    \toprule
    A. & Walk & D. & Sitting & G. & Wave hands &J. & Wiping \\
    B. & Run & E. & Empty room & H. & Clapping & K. & Squat \\
    C. & Jump & F. & Standing & I. & Lay down & L. & Stretching \\
    \bottomrule
  \end{tabular}
  \label{tab:activities}
  \vspace{-0.3cm}
\end{table}

\begin{figure}
  \centering
  \includegraphics[width=\columnwidth]{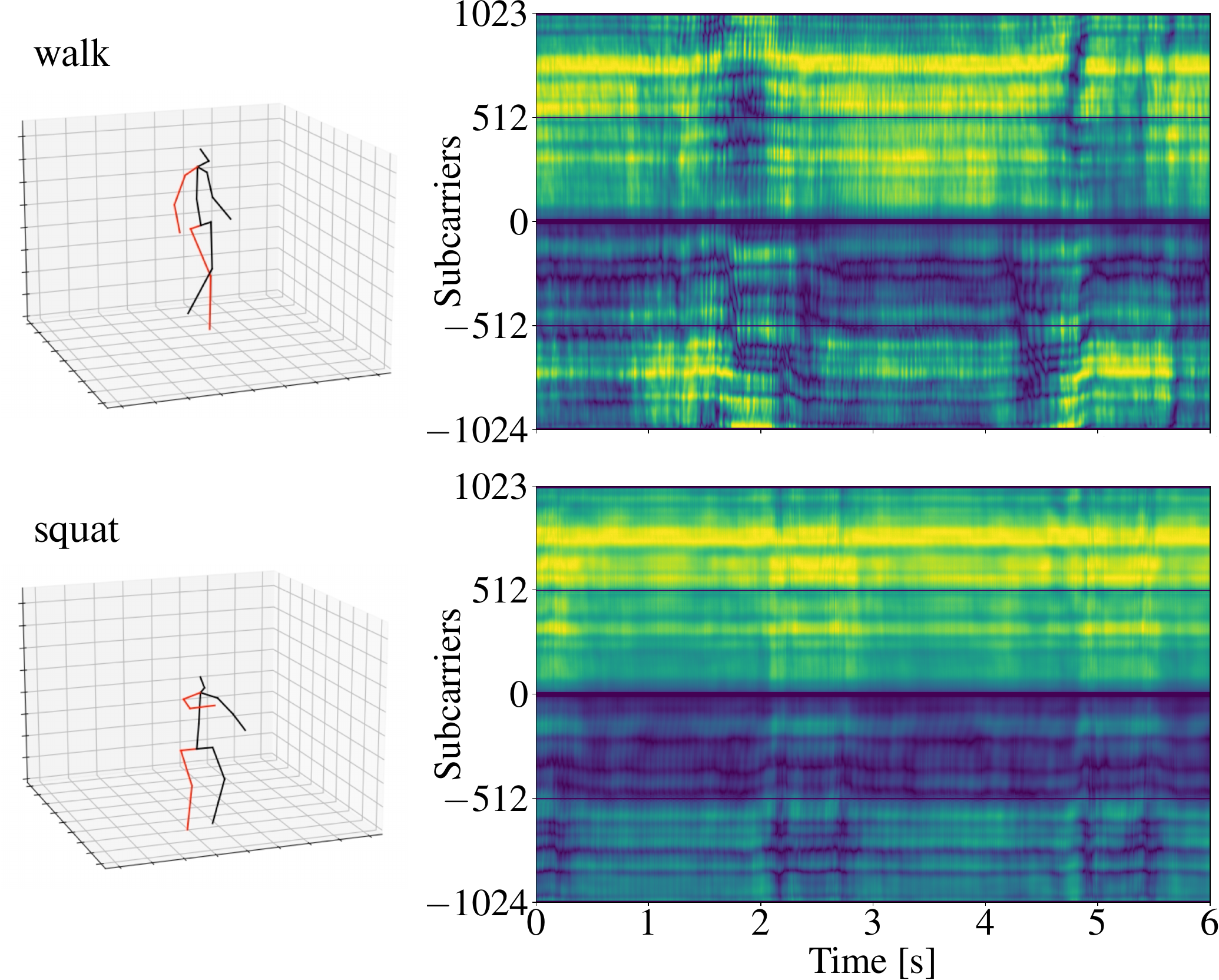}
  \caption{Samples of 3D models extracted from the video with the corresponding \acrshort{csi} for the activities \textit{walk} and \textit{squat}.
  }
  \label{fig:csi-skeleton}
\end{figure}

Another significant novelty is the recording of an anonymized video ground truth synchronized to the \gls{csi} data. Indeed, \gls{csi} data related to a particular activity is often provided without a satisfactory description of the process that generated the data, e.g., the frequency of the jumps or the exact motion while waving the hands.
To overcome this problem, we record the candidate with a camera while executing the target activities.
In this way, other researchers can work on our dataset in the future and know exactly how the activities are performed in every trace.
At this point, we must make a critical observation about the video recordings and the participants' anonymity.
The candidates consented to be video recorded during the experiments; however, their identity is concealed in the public dataset.
Indeed, the public ground truth consists of only a few key-points per frame (corresponding to the candidate's joints) and an anonymous 3D figure that is automatically modeled on the candidate's activities starting from the video trace, as shown in \cref{fig:csi-skeleton}.
To automatically reconstruct the 3D figure from video data we use VideoPose3D~\cite{pavllo20193d}, an open-source tool available online at \url{ https://github.com/facebookresearch/VideoPose3D}.

Finally, we briefly describe how we synchronized the \gls{csi} data to the video trace.
During every experiment, we first start recording the video using the webcam attached to the host laptop.
When the transmitter sends the first \wifi{} frame, an audiovisual signal is created and recorded in the video trace.
We assume this time also corresponds to the reception time of the same frame since propagation delays in indoor environments are negligible with respect to the standard video frame rate (30~Hz).
In the same way, an audiovisual signal is created for the last \wifi{} frame.
This method allows for simple synchronization of \gls{csi} data with a video trace without setting up a complex infrastructure, thus improving the replicability of the setup by anyone interested in expanding the dataset with common ground-truth data.

\subsection{CSI Data Structure}

Preliminary experiments showed that it is hard to achieve satisfactory results with raw \gls{csi} data, e.g., by training a \gls{cnn} with sequences of raw \gls{csi} taken from the same target activity.
Therefore, organizing the \gls{csi} in a data structure that is both informative and manageable is fundamental to support further processing.

We recall that every \gls{csi} \textit{data point} can be represented by a complex-valued matrix of size $W \times N$, where $W$ is the number of \gls{ofdm} subcarriers and $N$ is the number of receiving antennas (we omit multiple spatial streams in our discussion for simplicity).
We stack $T$ consecutive \gls{csi} data points along the ``time'' dimension to create a 3D tensor of size $T \times W \times N$.
This data structure enables easy retrieval of the \gls{csi} corresponding to a specific time instant, as well as studying the evolution of the \gls{csi} for single antennas (e.g., the \gls{csi} already shown in \cref{fig:csi-skeleton}) or along subsets of subcarriers.


\subsection{Activity Recognition Framework}

In the previous sections we described how to systematically collect and store large volumes of \gls{csi} data that are also synchronized to a video trace.
At this point, we only miss a system to benchmark how different features of the \wifi{} signal (i.e., bandwidth, \gls{ofdm} subcarrier spacing, and number of antennas) affect \gls{csi} sensing performance.

In this paper, we decided to employ SHARP\footnote{See \url{https://github.com/signetlabdei/SHARP}}, a state-of-the-art open-source system targeting person- and environment-independent activity recognition.
SHARP is a \textit{supervised} \gls{dl} framework that solves a \textit{classification} problem: it takes in input sequences of \gls{csi} and outputs an activity chosen from the set of possible activities on which it was trained.
The idea at the core of SHARP is first to preprocess the raw \gls{csi} data to extract the so-called \textit{Doppler vector}.
The Doppler vector is a feature that should not depend on the surrounding environment---unlike the \gls{csi}---but only on the ``variations'' across different \gls{csi} and, thus, on the specific activity being monitored.
Describing the SHARP system in detail is beyond the scope of this paper, and we redirect the interested reader to the original paper~\cite{meneghello2022} for further details about its theory and implementation.
However, we recall that SHARP has been designed to work with 80-MHz 802.11ac transmissions; hence, we had to customize some of the preprocessing operations to take into account the different number of subcarriers and the reduced subcarrier spacing.

\section{Experimental Results}

We finally present and discuss the results of our investigation.
We want to stress again that our primary research objective is to systematically investigate how different parameters of \wifi{} communications impact sensing performance.
We do not propose a novel sensing algorithm; instead, we study the performance of one of the latest \wifi{} sensing systems on an independent dataset we crafted for this task.

The results are subject to a broad spectrum of analysis and interpretation work, so large that the amount of data to look at quickly becomes overwhelming.
For this reason, and due to space constraints, we present only a few possible perspectives on the outcomes of our analysis and show the results that offer the most intriguing insights on fundamental questions about \gls{csi}-based sensing.

\subsection{Same Person, Same Environment (\texttt{PE})}

We start our analysis by presenting the results obtained when the system is trained with \gls{csi} data of one person in one environment and then tested using different data from the same person in the same environment.
These results represent a baseline and can later be used to study how the system generalizes to different subjects and environments.
For shorthand, we label this experiment \texttt{PE} (Person-Environment).

\begin{figure}
  \centering
  \includegraphics[width=0.96\columnwidth]{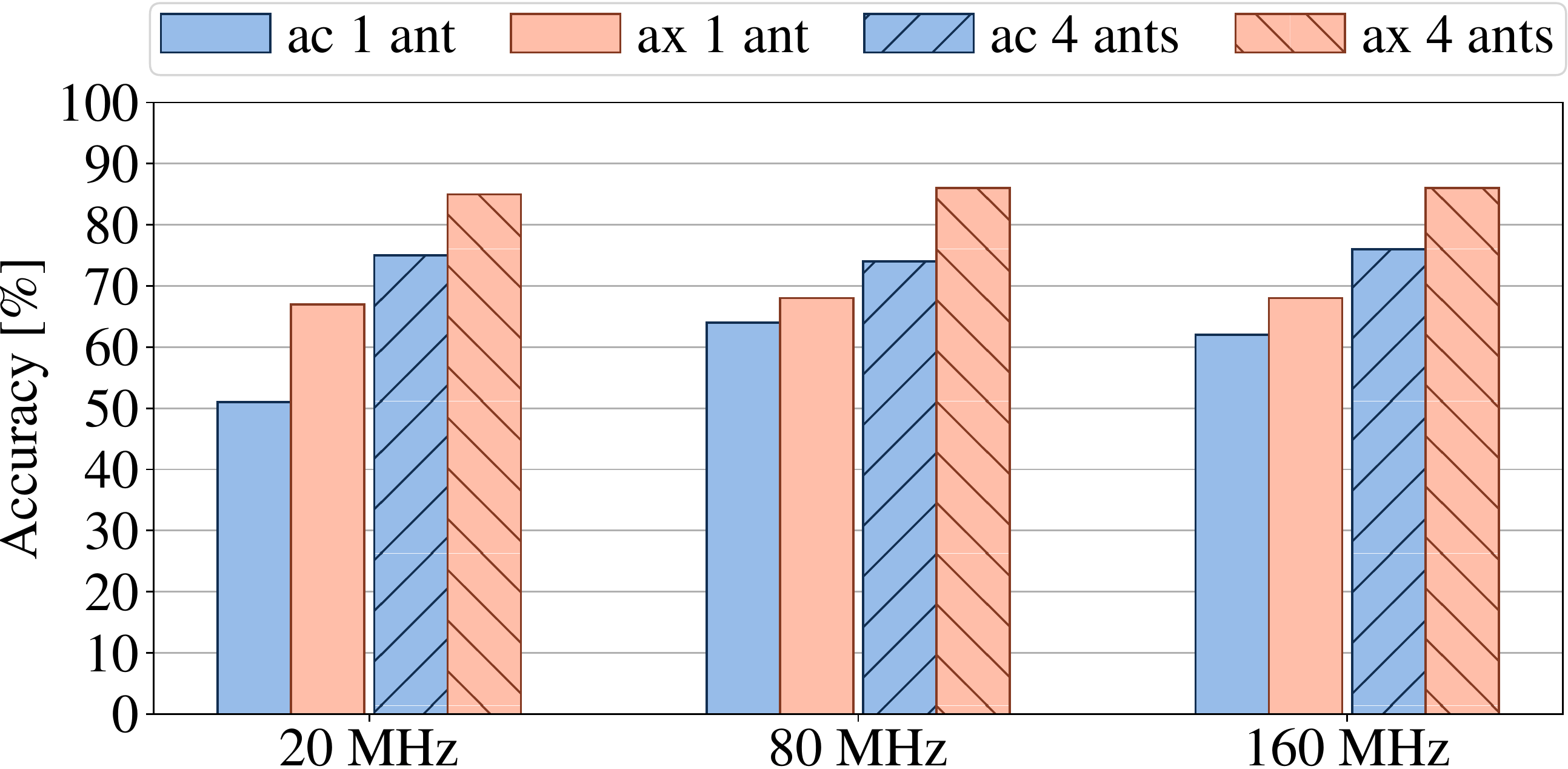}
  \caption{Average classification accuracy for 5 activities in the \texttt{PE} experiment. We consider different bandwidths using 802.11ac and 802.11ax systems with 1 and 4 antennas.}
  \label{fig:accuracy_pe_5classes}
  \vspace{-0.3cm}
\end{figure}

In \cref{fig:accuracy_pe_5classes}, we show the average \texttt{PE} accuracy (on scenarios S1 to S7 in \cref{tab:scenarios}) for the same five target classes that were considered in the original SHARP work (labeled from A to E in \cref{tab:activities}).
We can already take a few interesting insights from these results.
First, we notice that once we fix the bandwidth, the classification accuracy increases if we increase the spectral resolution and the number of receiving antennas.
The effect is evident regardless of the bandwidth, as the classification accuracy increases by 30 to 40 percent when moving from a single-antenna 802.11ac system to a four-antenna 802.11ax one.
Second, it is apparent that the \gls{csi} bandwidth does not play a primary role in the considered \gls{csi} sensing application, as the improvement from 20~MHz to 80~MHz and 160~MHz is marginal.
We guess that even if a wider bandwidth helps to characterize better the communication channel, the main features of human motion are already detected when considering smaller bandwidths, and their recognition does not improve as the bandwidth increases.

\begin{figure}
  \centering
  \begin{subfigure}{0.47\columnwidth}
    \centering 
    \includegraphics[width=\columnwidth]{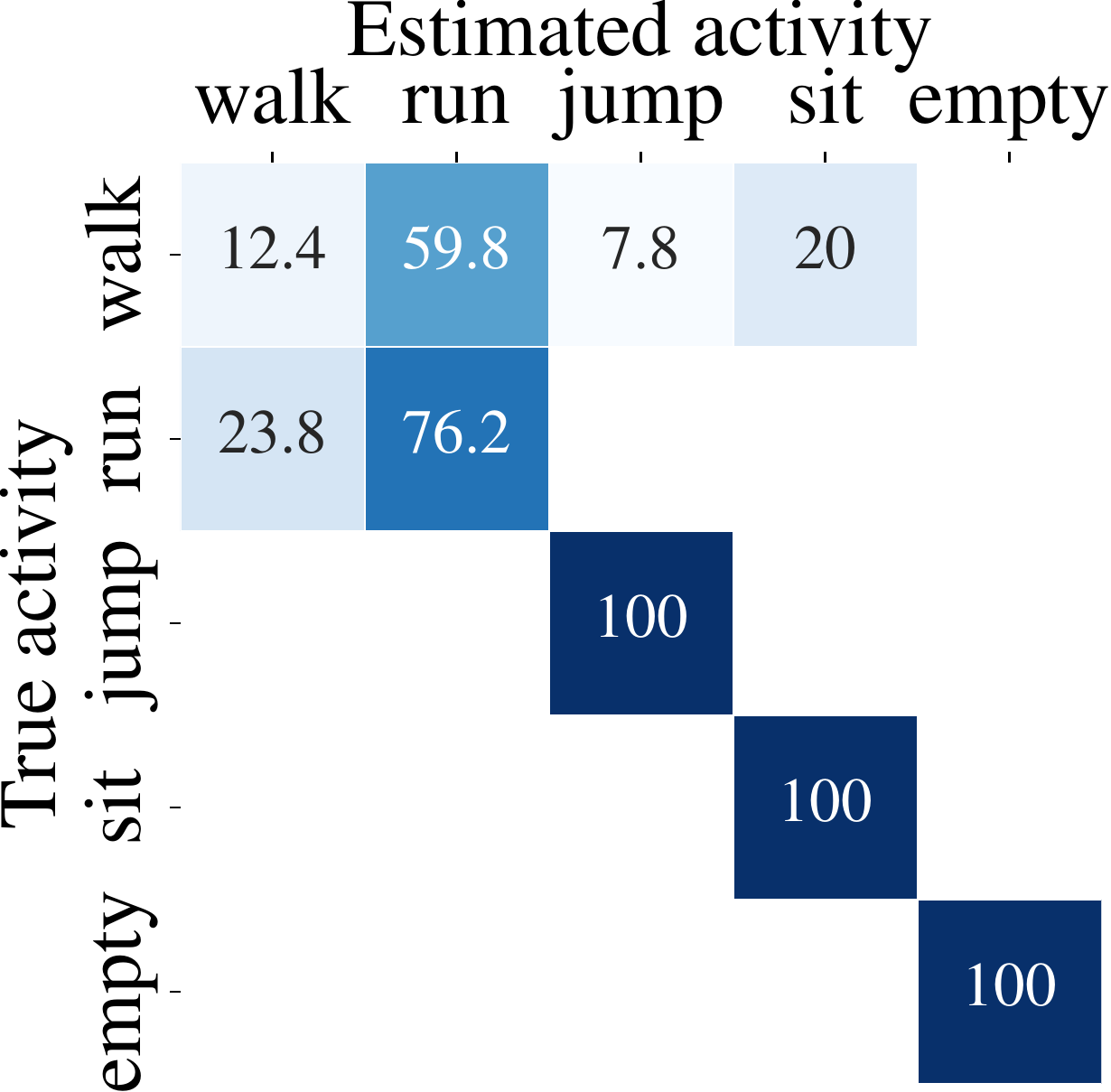}
    \caption{80 MHz, 802.11ac, 4 ants}
  \end{subfigure}
  ~
  \begin{subfigure}{0.47\columnwidth}
    \centering 
    \includegraphics[width=\columnwidth]{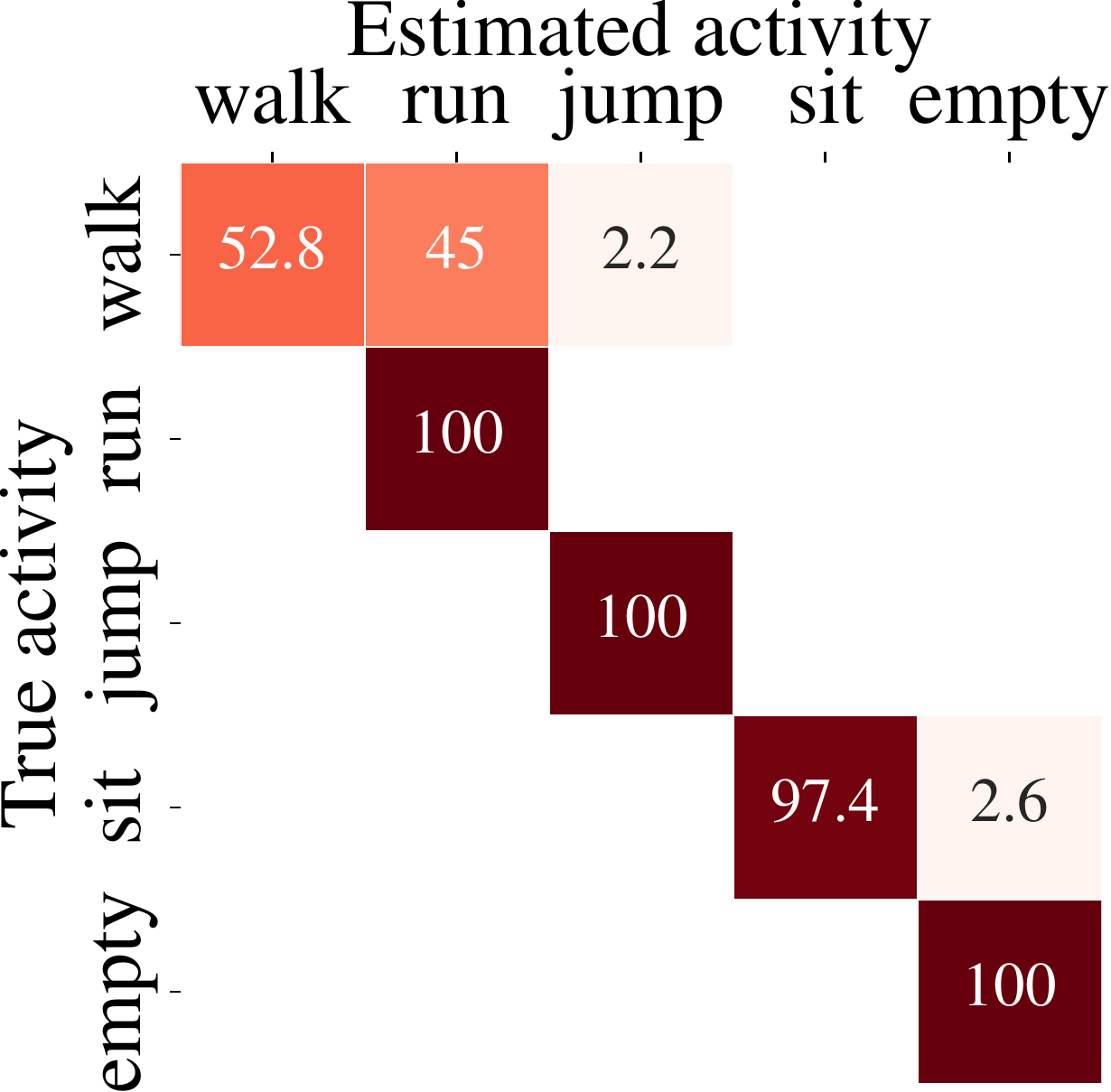}
    \caption{160 MHz, 802.11ax, 4 ants}
  \end{subfigure}
  \caption{Confusion matrices for 5 classes in the \texttt{PE} experiment. The values indicate percentages of true samples.}
  \label{fig:confmat_pe_5classes}
  \vspace{-0.2cm}
\end{figure}

There is another critical thing to notice in \cref{fig:accuracy_pe_5classes}.
We recall that the system achieved 100\% accuracy in the original SHARP paper, while here it stops at almost 90\% accuracy in the best case.
To further investigate this discrepancy, we show the confusion matrices for two different experiments in \cref{fig:confmat_pe_5classes}.
The confusion matrices clearly show a high misclassification rate between \textit{walk} and \textit{run}, while the other activities are perfectly detected.
This is reassuring in some sense because it is often hard to discern between these two activities, especially in indoor environments.
It is worth noticing that if we join together the classes \textit{walk} and \textit{run} (e.g., in a single class \textit{move}), then the classification accuracy touches 100\% for our dataset too.
However, this consideration also raises an important question about how \gls{csi} sensing models can generalize across different settings and repetitions of the same experiment.

\begin{figure}
  \centering
  \includegraphics[width=0.96\columnwidth]{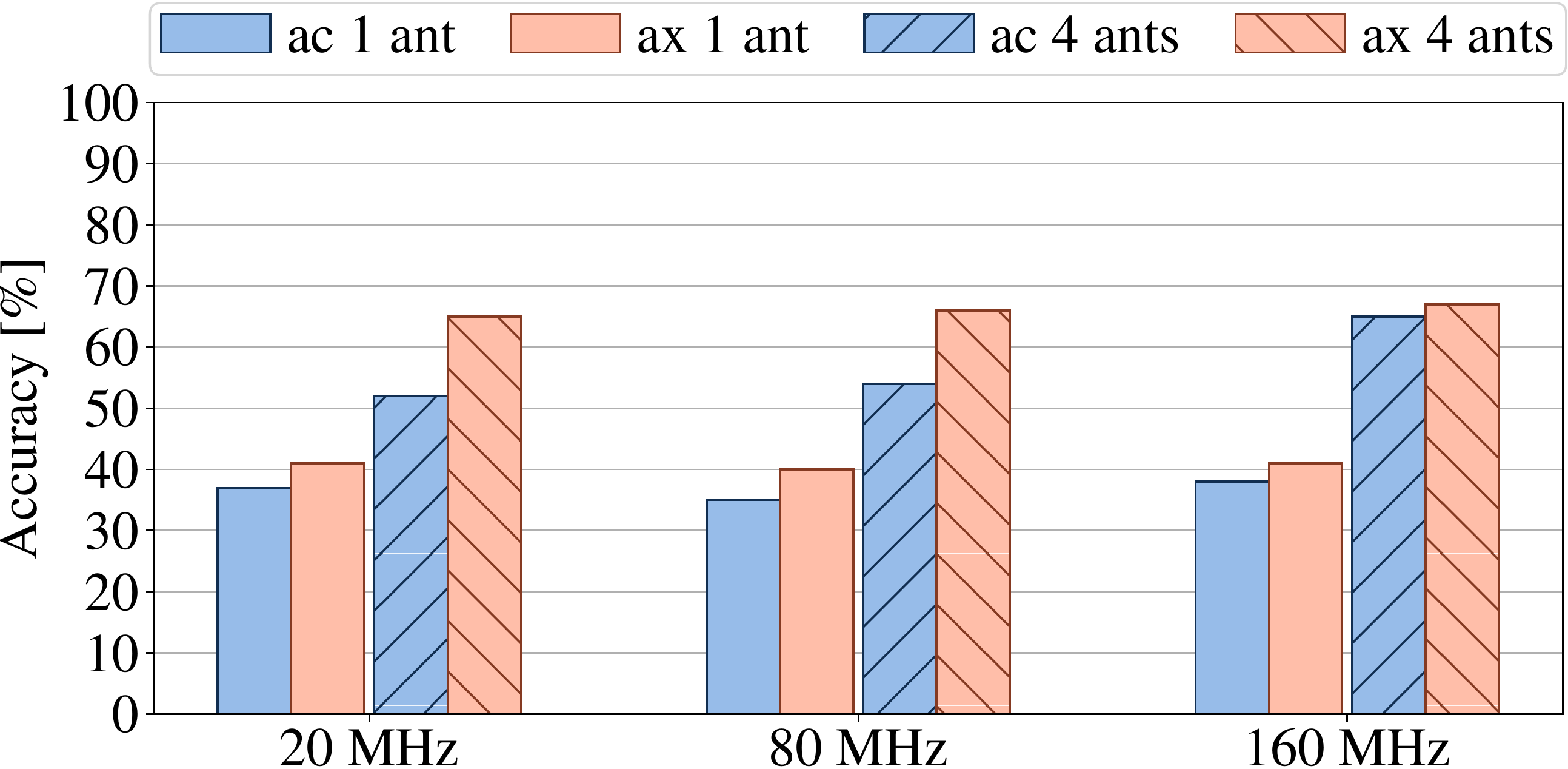}
  \caption{Average classification accuracy for 12 activities in the \texttt{PE} experiment. We consider different bandwidths using 802.11ac and 802.11ax systems with 1 and 4 antennas.}
  \label{fig:accuracy_pe_12classes}
  \vspace{-0.3cm}
\end{figure}

Another interesting question regards the performance of the system when we consider more activities.
In \cref{fig:accuracy_pe_12classes}, we show the accuracy in the \texttt{PE} experiment for 12 different activities.
As expected, the overall accuracy drops when we enlarge the set of target activities; however, we notice that with 802.11ax transmissions and four antennas, the average accuracy stays above 60\% (we point out that a random choice would be almost 8\%).
The results reproduce almost exactly the behavior we have already commented on for the \texttt{PE} experiment with 5 classes, except for the 160-MHz 802.11ac transmissions with 4 antennas, for which the larger bandwidth seems to correspond to a slight increase in the average accuracy.

\begin{figure}
  \centering
  \begin{subfigure}{0.47\columnwidth}
    \centering 
    \includegraphics[width=\columnwidth]{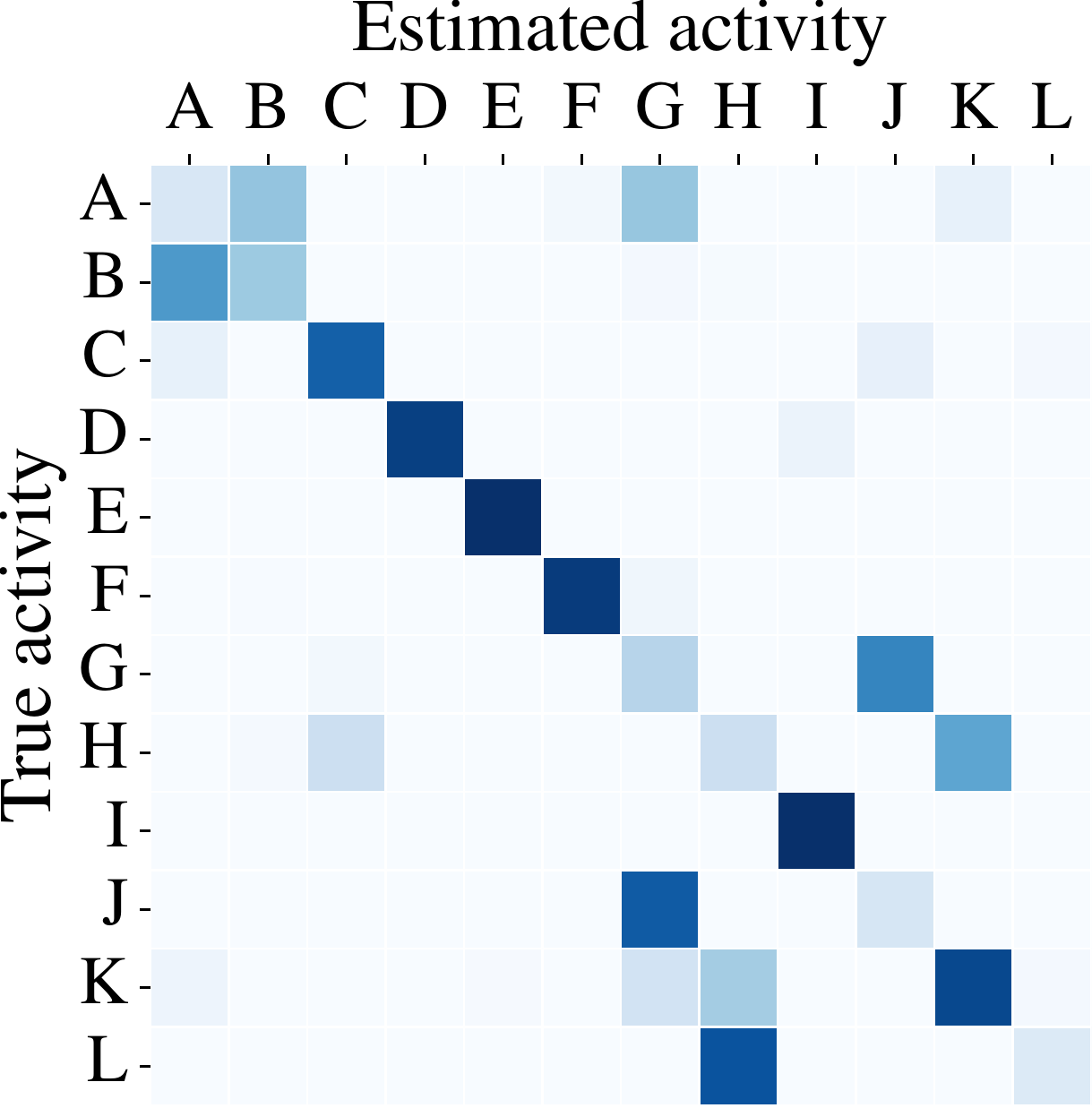}
    \caption{80 MHz, 802.11ac, 4 ants}
  \end{subfigure}
  ~
  \begin{subfigure}{0.47\columnwidth}
    \centering 
    \includegraphics[width=\columnwidth]{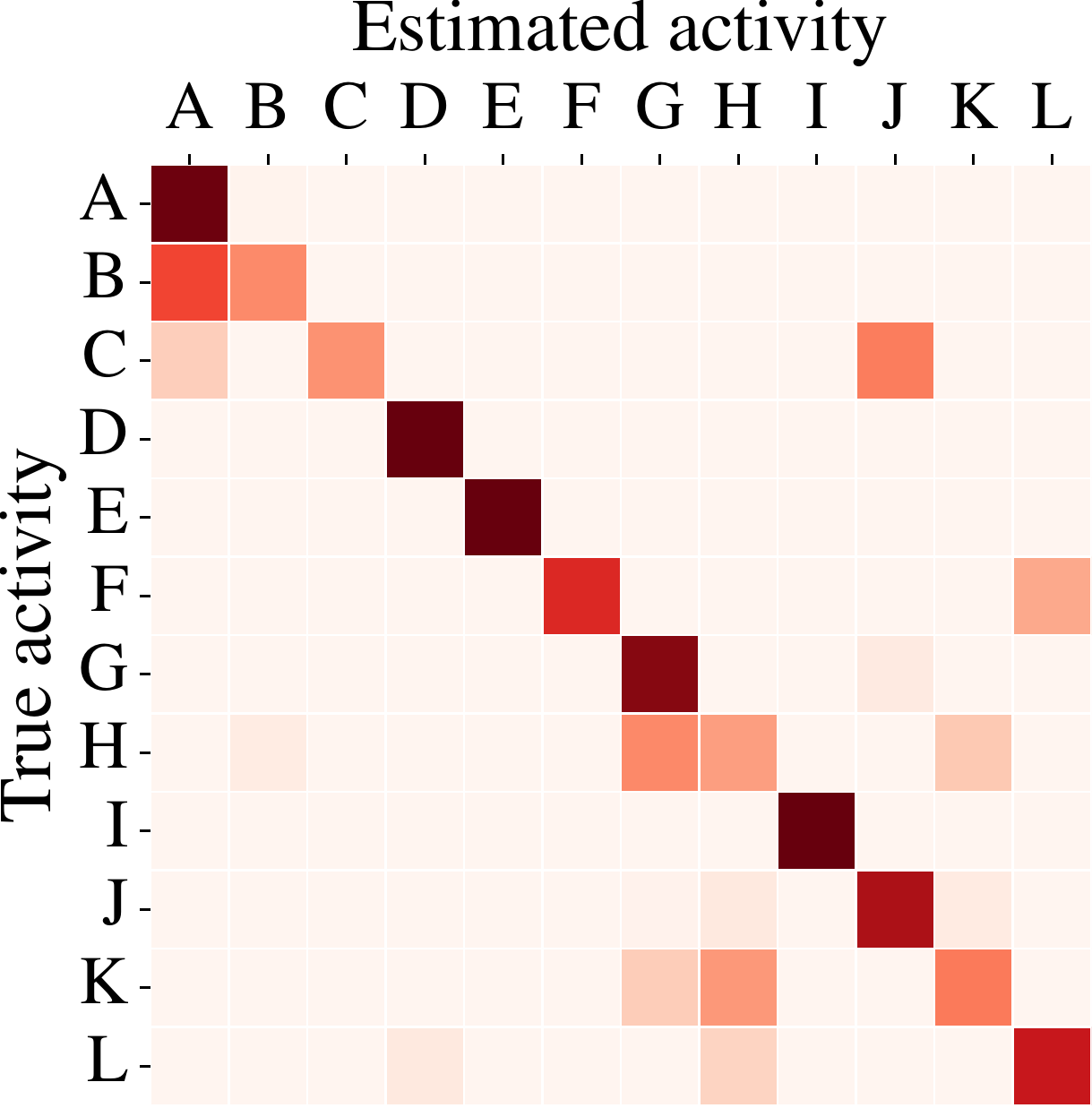}
    \caption{160 MHz, 802.11ax, 4 ants}
  \end{subfigure}
  \caption{Confusion matrices for 12 classes in the \texttt{PE} experiment. Darker colors mean higher values.}
  \label{fig:confmat_pe_12classes}
\end{figure}

We report two confusion matrices for this experiment in \cref{fig:confmat_pe_12classes}.
We observe that also in this case there is substantial confusion between the activities \textit{walk} and \textit{run}.
Moreover, we notice an interesting misclassification trend between pairs of activities such as \textit{wave hands} and \textit{wiping}, which is reasonable, or \textit{clapping} and \textit{squat}, which instead is a bit more obscure.

\begin{figure}
  \centering
  \includegraphics[width=0.96\columnwidth]{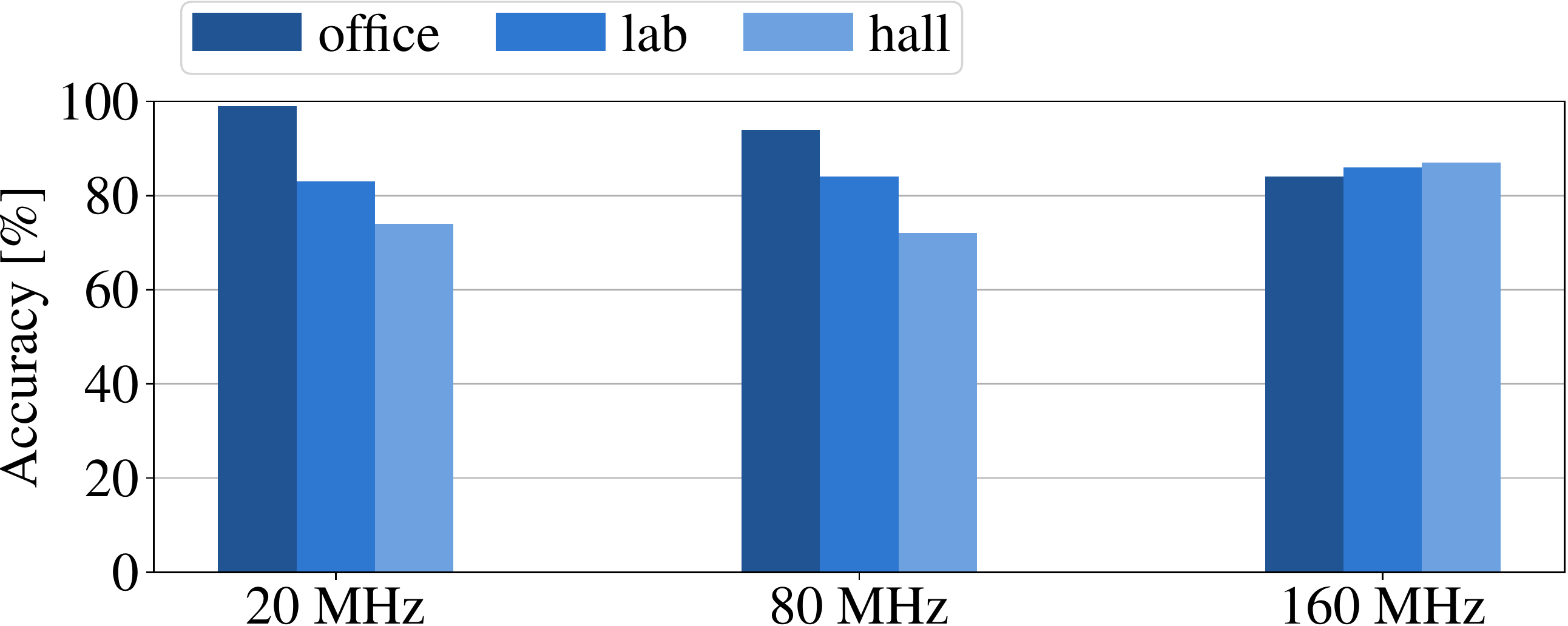}
  \caption{Classification accuracy for 5 activities in different environments using 802.11ax with 4 antennas.}
  \label{fig:accuracy_rooms}
  \vspace{-0.3cm}
\end{figure}

Finally, as we ran \texttt{PE} experiments in three different environments, it is interesting to ask ourselves whether the sensing performance is the same everywhere.
In \cref{fig:accuracy_rooms}, we show the classification accuracy obtained in different environments for 802.11ax transmissions with four receiving antennas and different bandwidths.
While the average accuracy remains the same across the different bandwidths (as we have already acknowledged in \cref{fig:accuracy_pe_5classes}), we can clearly distinguish other trends that depend on the environment.
In the office, the smallest of the three environments, the accuracy is higher when working with 20~MHz transmissions and decreases as the bandwidth grows.
We see the opposite trend in the hall, which is also the largest of the three environments, while in the lab the overall accuracy remains mostly unchanged.
This behavior confirms that the interference pattern created by different activities is indeed intertwined with the effect of the surrounding environment and might represent a problem when trying to generalize to different environments.

\subsection{Different person, same environment (\texttt{XP})}

We have presented the results obtained by training and testing the \wifi{} sensing system on \gls{csi} data collected from the same person in a single environment.
Now, we focus our attention on a different experiment, in which we test the ability of the system to work with activities performed by a subject different from the one on which the specific model was trained.
We label this kind of experiment as \texttt{XP}.

\begin{figure}
  \centering
  \includegraphics[width=0.96\columnwidth]{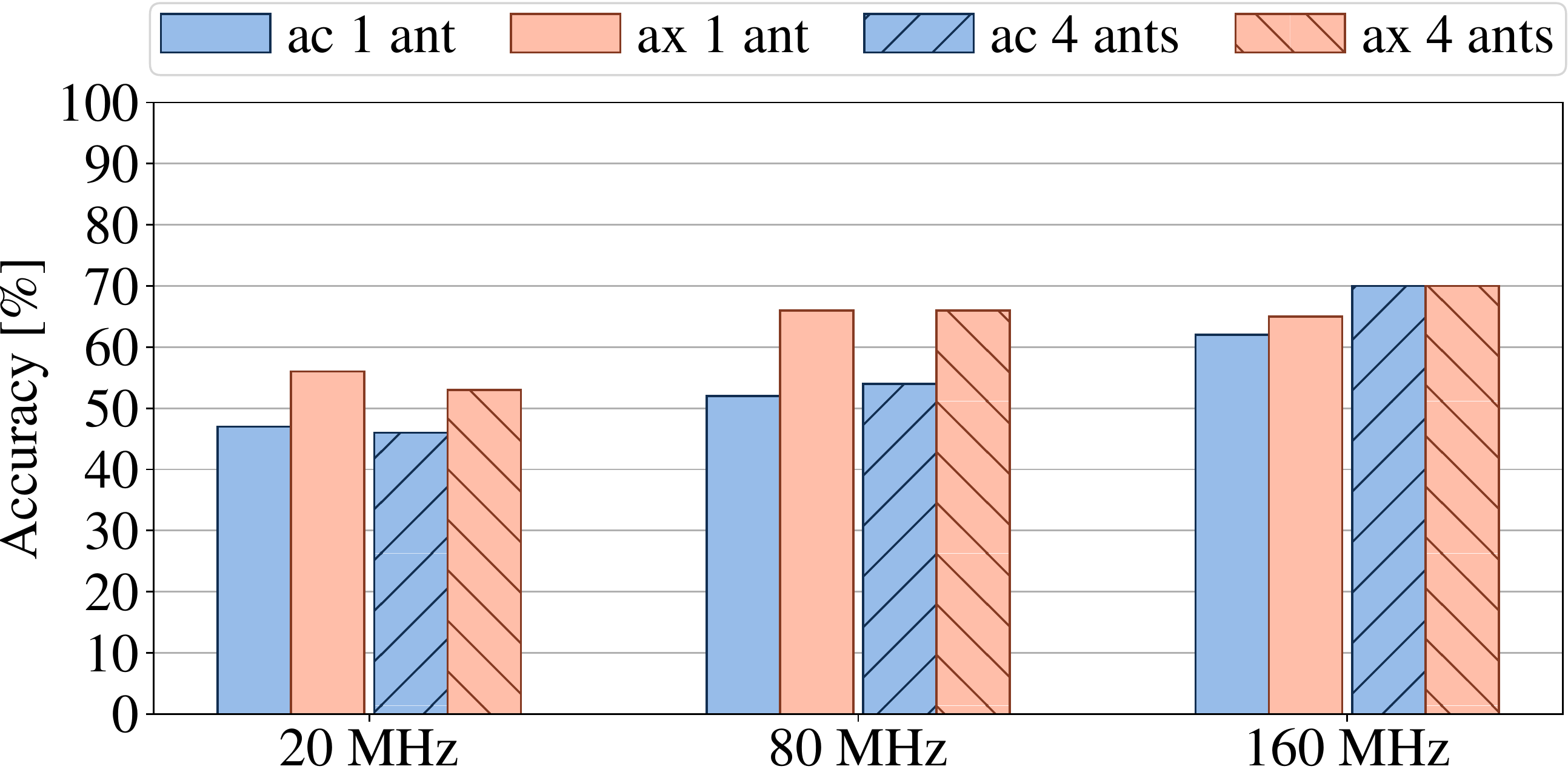}
  \caption{Average classification accuracy for 5 activities in the \texttt{XP} experiment. We consider different bandwidths using 802.11ac and 802.11ax systems with 1 and 4 antennas.}
  \label{fig:accuracy_xp_5classes}
  \vspace{-0.4cm}
\end{figure}

In \cref{fig:accuracy_xp_5classes}, we show the results obtained for the \texttt{XP} experiment over the same five classes we considered before.
We notice two main differences with respect to the \texttt{PE} results.
First, we see that, in this case, using more antennas does not improve classification accuracy.
While this might be confusing at first, this result seems to indicate that the trained model is indeed \textit{overfitting} on the training dataset.
Second, in the \texttt{XP} experiment we notice a weak correlation between the accuracy and the \gls{csi} bandwidth.
Unlike in the \texttt{PE} experiment, this result seems in line with the basic intuition that ``more data, the better.''
We suspect that a larger bandwidth helps the sensing system to detect similar features when two different users perform the same gesture.

In this case, the outcome of our experiment differs from the one reported in the original SHARP paper, which reports high accuracy for \texttt{XP} experiments.
However, we want to highlight the fact that people might perform the same gesture in different ways, e.g., with different speeds and periodicity.
This observation might undermine the assumption that different subjects cause similar variations in the \gls{csi} when performing the same activity.
We believe that in this case the video ground truth will help interpret patterns in the \gls{csi} data and differences between different realizations of the same gestures.

We noticed that the performance of the system is almost equivalent in the \texttt{XP} experiment and in other experiments in which we train and test the system on the same person in the same environment after some time has passed (i.e., training with dataset S1 and testing with dataset S4 and S5, with reference to \cref{tab:scenarios}).
This means that even if the system still works correctly in most cases, it is not entirely independent from the channel's current state, which slightly impairs the activity recognition performance.

\subsection{Same person, different environment (\texttt{XE})}

The last part of our study focuses on analyzing the sensing performance when the system is provided with \gls{csi} data of the same person for which it was trained, but the activities are performed in a different environment.
This is the most challenging scenario, and we label these experiments \texttt{XE} to recall the cross-environment nature.

Unfortunately, in our experiments, SHARP failed to generalize to new environments.
In the best cases, the system did only slightly better than a random choice between the target activities.
On the one hand, this result is quite expected because there is no guarantee that the same activities will have the same effect on the physical channel in different environments.
On the other hand, however, this result contradicts the environment-independent nature of the SHARP framework.
One possible explanation for this inconsistency is that macroscopic differences in the training and testing environments (e.g., different room sizes and amounts of reflective clutter) can disrupt the performance of the sensing system.
In the original SHARP work, all the rooms had more or less the same size; however, in this work, we chose different classes of rooms to test the system's limits.
We believe this point raises fundamental questions about the generalization capabilities of current \gls{csi} sensing models, opening several new issues to address in this research field and encouraging authors to work collaboratively on datasets as heterogeneous as possible.

\section{Concluding Remarks}

This paper presents the first full-fledged investigation into \gls{csi}-based sensing with \wifi{}~6 signals.
We have set up an extensive data collection campaign involving 3 human subjects and 3 indoor environments, measuring the impact of 12 activities on the \gls{csi} in 7 different scenarios.
We have collected \gls{csi} measurements using 3 nodes with 4 receiving antennas each and up to 2048 subcarriers per receiving antenna, leading to up to 8192 data points per \gls{csi}.
We have created an anonymized ground truth based on a video recording of the subjects performing every activity.
We have leveraged our dataset to dissect the impact that different \wifi{}~6 features (i.e., available bandwidth, subcarrier spacing, and multiple receiving antennas) have on the classification performance.
We have leveraged a state-of-the-art \gls{dl} algorithm for benchmarking purposes and evaluated its generalization performance when samples collected in different environments and with other subjects are fed to the \gls{dl} model.

\subsection*{\textbf{Significance and Impact of Results}}

We studied how different features of the \wifi{} signals impact the performance of a state-of-the-art sensing framework.
Our experimental results indicate that despite ten years of innovative research in the field of \gls{csi}-based \wifi{} sensing, there is still much more to understand and discover about this topic.
While our results verify the main findings of the research community, they also suggest that achieving reliable generalization to new, unseen situations and environments is a difficult task.
Beyond the results discussed in this paper, we firmly believe that the community should conduct further investigations into CSI sensing.
For this reason, and to allow full replicability of our results, we release to the research community our labeled CSI dataset for about 80 GB of data.

\section{Acknowledgement of Support and Disclaimer} 
\noindent This material is based upon work supported by the National Science Foundation under Grant No. CNS-2134973 and CNS-2120447, and co-funded/supported by the German Research Foundation (DFG) in the Collaborative Research Center (SFB) 1053 MAKI. Any opinions, findings, and conclusions or recommendations expressed in this material are those of the author(s) and do not necessarily reflect the views of the National Science Foundation.

\small
\bibliographystyle{ieeetr}
\bibliography{references}

\end{document}